\documentclass[12pt]{article}
\usepackage{arxiv}
\usepackage{tmath}
\usepackage{apacite}
\usepackage{graphicx}
\usepackage{amsmath, amssymb}
\usepackage[T1]{fontenc}    
\usepackage{amsfonts}       
\usepackage{nicefrac}       
\usepackage{microtype}      
\usepackage{lineno}

\mathchardef\mhyphen="2D

\title{A ``Trap-Release-Amplify'' Model of Chorus Waves}
\author{
  Xin Tao\thanks{Correspondence to: \texttt{xtao@ustc.edu.cn}.} \\
  Department of Geophysics and Planetary Sciences, University of Science and Technology of China, Hefei, China \\
  \And
  Fulvio Zonca \\
  Center for Nonlinear Plasma Science and C.R. ENEA Frascati, CP 65-00044 Frascati, Italy
  \AND
  Liu Chen \\
  Institute of Fusion Theory and Simulation, Department of Physics, Zhejiang University, Hangzhou, China
}

\begin{document}
\maketitle




  


\begin{abstract}
Whistler mode chorus waves are quasi-coherent electromagnetic emissions with frequency chirping.  Various models have been proposed to understand the chirping mechanism, which is a long-standing problem in space plasmas. Based on analysis of effective wave growth rate and electron phase space dynamics in a self-consistent particle simulation, we propose here a phenomenological model called the ``Trap-Release-Amplify'' (TaRA) model for chorus. In this model, phase space structures of correlated electrons are formed by nonlinear wave particle interactions, which mainly occur in the downstream. When released from the wave packet in the upstream, these electrons selectively amplify new emissions which satisfy the phase-locking condition to maximize wave power transfer, leading to frequency chirping. The phase-locking condition at the release point gives a frequency chirping rate that is fully consistent with the one by Helliwell in case of a nonuniform background magnetic field. The nonlinear wave particle interaction part of the TaRA model results in a chirping rate that is proportional to wave amplitude, a conclusion originally reached by Vomvoridis et al. Therefore, the TaRA model unifies two different results from seemingly unrelated studies. Furthermore, the TaRA model naturally explains fine structures of chorus waves, including subpackets and bandwidth, and their evolution through dynamics of phase-trapped electrons. Finally, we suggest that this model could be applied to explain other related phenomena, including frequency chirping of chorus in a uniform background magnetic field and of electromagnetic ion cyclotron waves in the magnetosphere.        
\end{abstract}

\section{Introduction}
Whistler mode chorus is one of the most intense naturally occurring electromagnetic emissions in planetary magnetospheres. These waves are practically important because they play key roles in energetic electron dynamics in the inner magnetosphere through resonant wave particle interactions, such as accelerating radiation belt electrons in the recovery phase of geomagnetic storms \cite{Horne2005b,Thorne2013b,Reeves2013}, or scattering plasmasheet electrons into the atmosphere to form diffuse and pulsating aurora \cite{Thorne2010,Nishimura2010}. Chorus waves are also scientifically interesting because they consist of narrowband quasi-coherent emissions with frequency chirping, which could occur in either upward (rising-tone) or downward (falling-tone) directions \cite{Tsurutani1974,Burtis1976}. The frequency chirping has also been found in other wave modes, such as electromagnetic ion cyclotron waves in the magnetosphere \cite{Pickett2010} or Alfv\'{e}n waves in fusion plasmas \cite{Heidbrink1995}, suggesting its universal presence. Correspondingly, understanding the chirping of chorus could be potentially beneficial to understanding a wide range of phenomena in both space and fusion plasmas, and is also the focus of this paper.  

The chirping mechanism of chorus waves has been under intensive research and debate since 1960s \cite{Helliwell1967,Sudan1971,Nunn1974,Vomvoridis1982,Sagdeev1985,Trakhtengerts1995,Omura2008,Omura2011,Demekhov2011,Zonca2017}. \citeA{Helliwell1967} proposed a phenomenological model to explain chirping by assuming a ``consistent-wave'' condition, in which the spatial variation of electron cyclotron frequency due to an inhomogeneous background magnetic field is matched by the Doppler shifted wave frequency to maximize wave power transfer. The chirping rate proposed in the model is proportional to the background magnetic field inhomogeneity and has been shown to be consistent with observation \cite{Tao2012b} and self-consistent particle-in-cell (PIC) simulations \cite{Tao2014b}. \citeA{Nunn1974} derived wave kinetic equations for narrowband emissions, and calculated the nonlinear resonant current due to phase-trapped electrons. Based on this analysis, the author developed a reduced numerical model called Vlasov Hybrid model, which is capable of simulating both rising tone and falling tone emissions \cite{Nunn1990,Nunn1997}. Assuming that chirping exists, \citeA{Vomvoridis1982} suggested that the chirping rate is proportional to wave amplitude, based on the maximization of wave amplification due to nonlinear wave particle interactions. This relation has also been obtained by \citeA{Omura2008} and \citeA{Zonca2017} with different methods, and used by the backward wave oscillator model of chorus \cite{Trakhtengerts1995} to obtain frequency chirping rate from the derived wave amplitude. It has also been verified directly by PIC simulations \cite{Katoh2013,Hikishima2009,Tao2017b} and observations \cite{Cully2011}. \citeA{Omura2011} proposed a sequential triggering model where chirping is caused by the nonlinear frequency shift due to the nonlinear current parallel to the wave magnetic field ($\delta j_B$). \citeA{Zonca2017} proposed a self-consistent theoretical framework, in which chirping is due to the nonlinear excitation of a narrowband spectrum out of a broad and dense background whistler wave modes \cite<see also,>{Zonca2015,Chen2016, Zonca2021}.  

Despite success of different chorus wave models in different aspects, various questions remain to be elucidated about the chirping process. First, if chirping is caused by the background magnetic field nonuniformity as in \citeA{Helliwell1967} or \citeA{Sudan1971}, how to explain chirping in a uniform background magnetic field as demonstrated by first-principle PIC simulations \cite{Wu2020b} or the BWO model simulation \cite{Demekhov2008}? Second, the chirping rate has been shown to be a function of either the background magnetic field inhomogeneity \cite{Helliwell1967} or the wave magnetic field \cite{Vomvoridis1982,Omura2008,Zonca2017}. These two estimates of the chirping rate are very different, but both have been verified by observations and simulations \cite{Tao2012b,Cully2011,Katoh2011a,Hikishima2009,Tao2017b}. Is it possible for both chirping rate equations to be correct? If so, how to properly relate one to the other and what is the cause of the difference? A better understanding of the chirping process of chorus is crucial to resolve these seemingly inconsistent results.  

In this work, based on detailed analysis of the spatial dependence of effective wave growth rate (Section \ref{sec:effect-growth-rate}) and electron phase space dynamics (Section \ref{sec:electron-phase-space}) in a PIC  simulation of chorus, and the theoretical framework of chorus by \citeA{Zonca2017}, we propose a ``Trap-Release-Amplify'' (TaRA) model to elucidate how and where chirping occurs (Section \ref{sec:trap-release-tar}). The model naturally yields both chirping rates of chorus from \citeA{Helliwell1967} and \citeA{Vomvoridis1982}. Detailed comparison of the TaRA model with models proposed by \citeA{Helliwell1967} and \citeA{Omura2011} is shown in Section \ref{sec:comp-with-prev}. Application of the TaRA model to explain chorus subpackets and instantaneous bandwidth is discussed in Section \ref{sec:fine-struct-chor}. We summarize our main results and discuss possible applications of the TaRA model to other chirping phenomena in Section \ref{sec:summary}.


\section{Simulation Setup}
\label{sec:simulation-setup}
To simulate chirping elements of chorus, we use a 1D spatial, 3D velocity code named DAWN \cite{Tao2014b}, originally developed after the 1D Electron Hybrid Model \cite{Katoh2007a} but later extended to use the nonlinear $\delta f$ method \cite{Parker1993,Hu1994} to reduce simulation noise \cite{Tao2017b}. Cold electrons are represented using linearized fluid equations, and hot electrons are modeled by PIC techniques with the nonlinear $\delta f$ method. The hot electron distribution is bi-Maxwellian with temperature anisotropy to provide the free energy for chorus excitation. Because whistler-mode chorus waves typically have a frequency much larger than the proton cyclotron frequency, ions are fixed. The background magnetic field is chosen to be of the parabolic form, $B=B_0 (1+\xi z^2)$. The flexible form of $B$ can represent a dipole field near the equator \cite{Helliwell1967,Nunn1990,Katoh2007a}, in which case the inhomogeneity parameter $\xi = 4.5/(L R_p)^2$ with $L$ the $L$-shell and $R_p$ the planet radius. It is also possible to investigate the controlling effects of magnetic field geometry on chorus chirping directions by changing the sign of $\xi$, as demonstrated by \citeA{Wu2020b}. Only parallel propagating whistler waves are allowed in the DAWN code. Reflecting and absorbing boundaries conditions are used for particles and waves, respectively. 

\begin{figure}
  \centering
  \includegraphics[width=\textwidth]{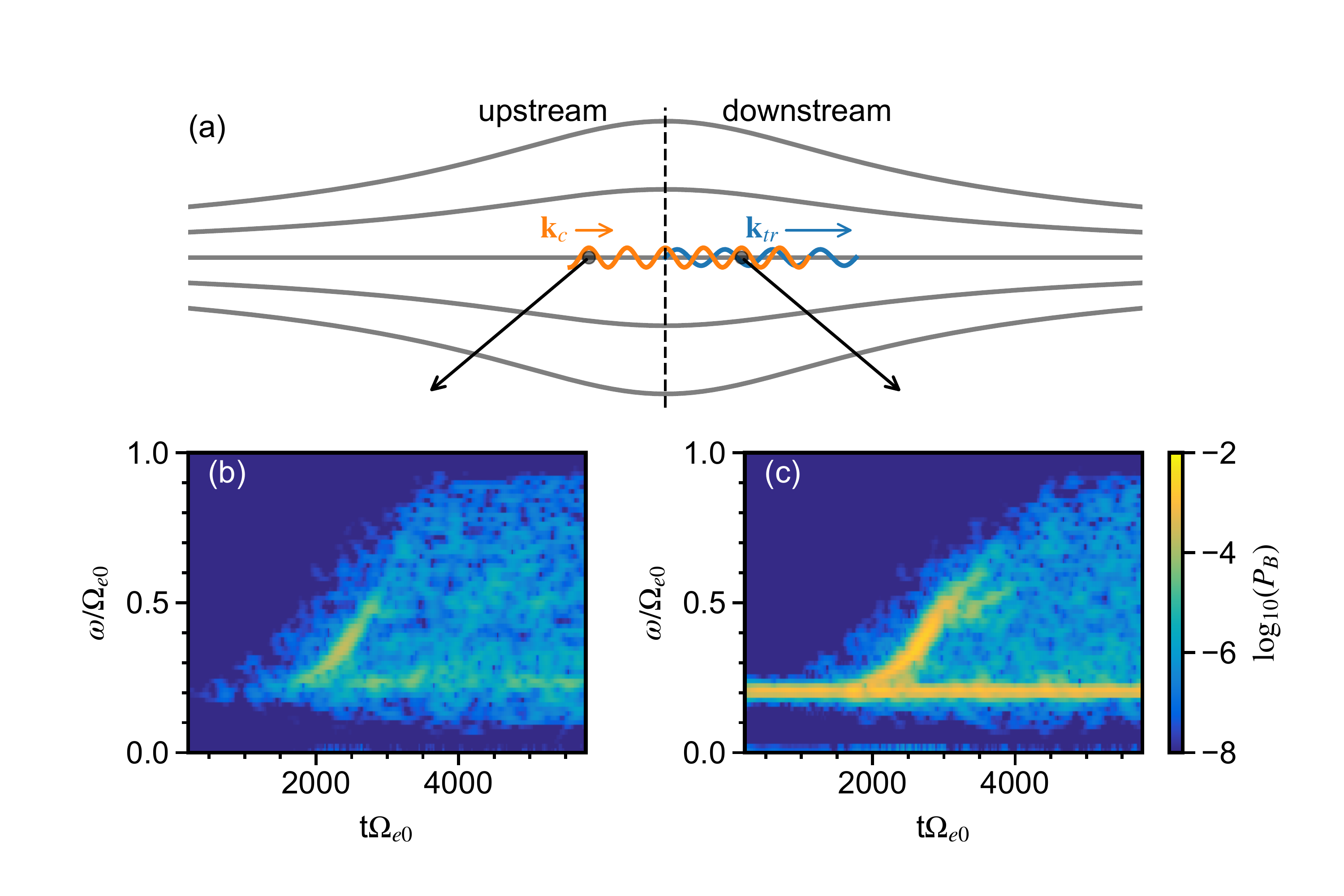}
  \caption{\label{fig:simulation_setup} (a) Illustration of the PIC simulation setup, and the definition of ``upstream'' and ``downstream'' regions. The blue waveform ($k_{tr}$) represents the unidirectional triggering wave used in the simulation, and the orange waveform ($k_c$) represents the generated chorus element. The vertical dashed line marks the equator ($z=0$). Example wave spectrograms showing chorus element in the upstream (b) and downstream (c). Color-coded is the wave magnetic field power spectral density.}
\end{figure}

Our simulation setup is illustrated in Figure \ref{fig:simulation_setup}a. A rightward propagating (in $+z$ direction) triggering wave, represented by $k_{tr}$, is launched from the equator. A chirping element of chorus, represented by $k_c$, is generated in the ``upstream'' region. This chorus element propagates in the same direction as the triggering wave, a result of cyclotron resonance between parallel propagating whistler waves and electrons. Here ``upstream'' and ``downstream'' regions are defined relative to the wave propagation direction. The reason we use a unidirectional triggering wave at the equator is that we would like to clearly demonstrate that the excitation of chorus occurs in the upstream region, as shown in Figure \ref{fig:simulation_setup}b and \ref{fig:simulation_setup}c. The generation of chorus element only in one direction also helps analyzing wave growth rate and electron phase space dynamics. Strictly speaking, this simulation is about triggered emissions; however, our conclusions in the paper apply equally well to spontaneous chirping of chorus, in which case, the most unstable linear mode plays the role of the triggering wave, as noted by several previous studies \cite<e.g.,>{Katoh2007a,Omura2008, Tao2020}. 

For our simulation, we use a time step $t\Omega_{e0}=0.02$ and a grid size $\Delta z = 0.05 d_e$ with $d_e \equiv c/\Omega_{e0}$ to accurately resolve the electron cyclotron motion and to satisfy the Courant condition. Here $\Omega_{e0}\equiv eB_0/mc $ is the equatorial electron cyclotron frequency, with $-e$ and $m$ the electron charge and mass, respectively, and $c$ being the speed of light in vacuum. Following previous studies \cite{Tao2014c,Tao2017a,Tao2017b}, the cold plasma frequency is chosen to be $\omega_{pe}=5\Omega_{e0}$, the ratio of hot to cold electron number density is $6\%$, and the background magnetic field inhomogeneity parameter $\xi = 2.155\times 10^{-5} d_e^{-2}$. In total, we use $6554$ cells with $2000$ particles per cell to reduce simulation noise. The parallel and perpendicular thermal velocities are $0.2\,c$ and $0.3\,c$, respectively. The corresponding temperature anisotropy is smaller than the one used by our previous studies \cite{Tao2014c, Tao2017a,Tao2017b}, so that no chirping elements can be generated without the triggering wave. Examples of chirping elements in the upstream and downstream regions can be found in Figure \ref{fig:simulation_setup}b and \ref{fig:simulation_setup}c.      

\section{Effective Growth Rate}
\label{sec:effect-growth-rate}
We estimate the effective growth rate of chorus waves to demonstrate its spatial dependence and to help identify the wave generation region.  

\subsection{Method of Estimation} 
This effective growth rate ($\gamma_\mathrm{eff}$) can be easily obtained from the wave kinetic equation
\begin{linenomath}
\begin{align}
  \pd{W}{t} + \pp{z}{\left(v_g W\right)} = 2\gamma_\mathrm{eff} W,
\end{align}
\end{linenomath}
as shown by \citeA{Zonca2017}, where $W$ is the wave energy density, and $v_g$ is the wave group velocity. It can also be obtained by noting that the time-averaged power transfer rate ($P$) is
\begin{linenomath}
\begin{align}
  \langle P \rangle  = -\frac{1}{2} \mathrm{Re}\left(\delta \vec{j}_h\cdot \delta \E^*\right),
\end{align}
\end{linenomath}
where $\langle \cdots \rangle$ represents averaging over fast wave oscillation period, $\delta \vec{j}_h$ and $\delta \E$ are complex notation of hot electron current density and wave electric field, respectively, with $\delta \E^*$ indicating the complex conjugate of $\delta \E$. The wave energy density, on the other hand, is given by \cite[pp 20]{Tao2020,Stix1992}
\begin{linenomath}
\begin{align}
  \label{eq:14}
  W = \frac{1}{16\pi} \frac{\omega}{v_g} \pd{D}{k} |\delta \E|^2 = -\frac{\omega}{16\pi}\pd{D}{\omega} |\delta \E|^2,
\end{align}
\end{linenomath}
where in the last equation, we have used that $v_g = - (\partial D/\partial k)/(\partial D/\partial \omega)$ with $k$ the wave number. Here
\begin{linenomath}
\begin{align}
  D = \frac{c^2k^2}{\omega^2}-\left[1-\frac{\omega_{pe}^2}{\omega(\Omega_e-\omega)}\right],
\end{align}
\end{linenomath}
is the local dispersion function of parallel propagating whistler waves with $\Omega_e=eB/mc$ the local electron cyclotron frequency. Correspondingly, the effective growth rate is given by
\begin{linenomath}
\begin{align}
  \label{eq:2}
  \gamma_\text{eff} = \frac{\langle P \rangle}{2W} = \frac{4\pi}{\omega}\left(\pd{D}{\omega}\right)^{-1} \frac{\mathrm{Re}\left(\delta \vec{j}_h\cdot \delta \E^*\right)}{|\delta \E|^2},  
\end{align}
\end{linenomath}
and it is a time dependent function of $\omega$ and $z$. We would like to remark that whether $\gamma_\text{eff}$ is nonlinear or not depends on the nonlinearity of $\delta \vec{j}_h$, or equivalently, the hot electron distribution function. We will now drop the subscript ``eff'' for simplicity.

As with other ways of obtaining growth rate in PIC type simulations, a reliable estimate depends on the signal-to-noise ratio. Therefore, as a consistency check, we also calculate the effective growth rate in a different way by using
\begin{linenomath}
\begin{align}
  \label{eq:3}
  W(z+\Delta z) = W(z) \exp\left[2 \int_z^{z+\Delta z} \left(\gamma'/v_g\right) \md z'\right] \approx W(z) \exp\left[2 \left(\gamma'/v_g\right) \Delta z\right].  
\end{align}
\end{linenomath}
Correspondingly,
\begin{linenomath}
\begin{align}
  \label{eq:13}
\gamma' = \frac{v_g}{2\Delta z}\ln\left[\frac{W(z+\Delta z)}{W(z)}\right].  
\end{align}
\end{linenomath}
A reliable estimate of effective growth rate would require that $\gamma$ and $\gamma'$ are consistent with each other. Otherwise, noise might dominate wave signal, and $\gamma$ (or $\gamma'$) cannot be used. 

\subsection{Variation of $\gamma$ along a ray path}
We investigate the spatial dependence of $\gamma$ and demonstrate generation of chorus in the upstream region by estimating $\gamma$ for a representative frequency, $\omega=0.34\Omega_{e0}$, along its ray path. This frequency is one of the frequencies from FFT and also about the frequency of maximum wave intensity, as indicated by the cross sign in Figure \ref{fig:zonca}b. We use $4096$ data points with $\Delta t=0.1 \Omega_{e0}^{-1}$ in all FFT calculations for wave spectrum and calculation of $\delta \vec{j}_h$ and $\delta \E$ used by Equation (\ref{eq:2}). 

\begin{figure}
  \centering
  \includegraphics[width=\textwidth]{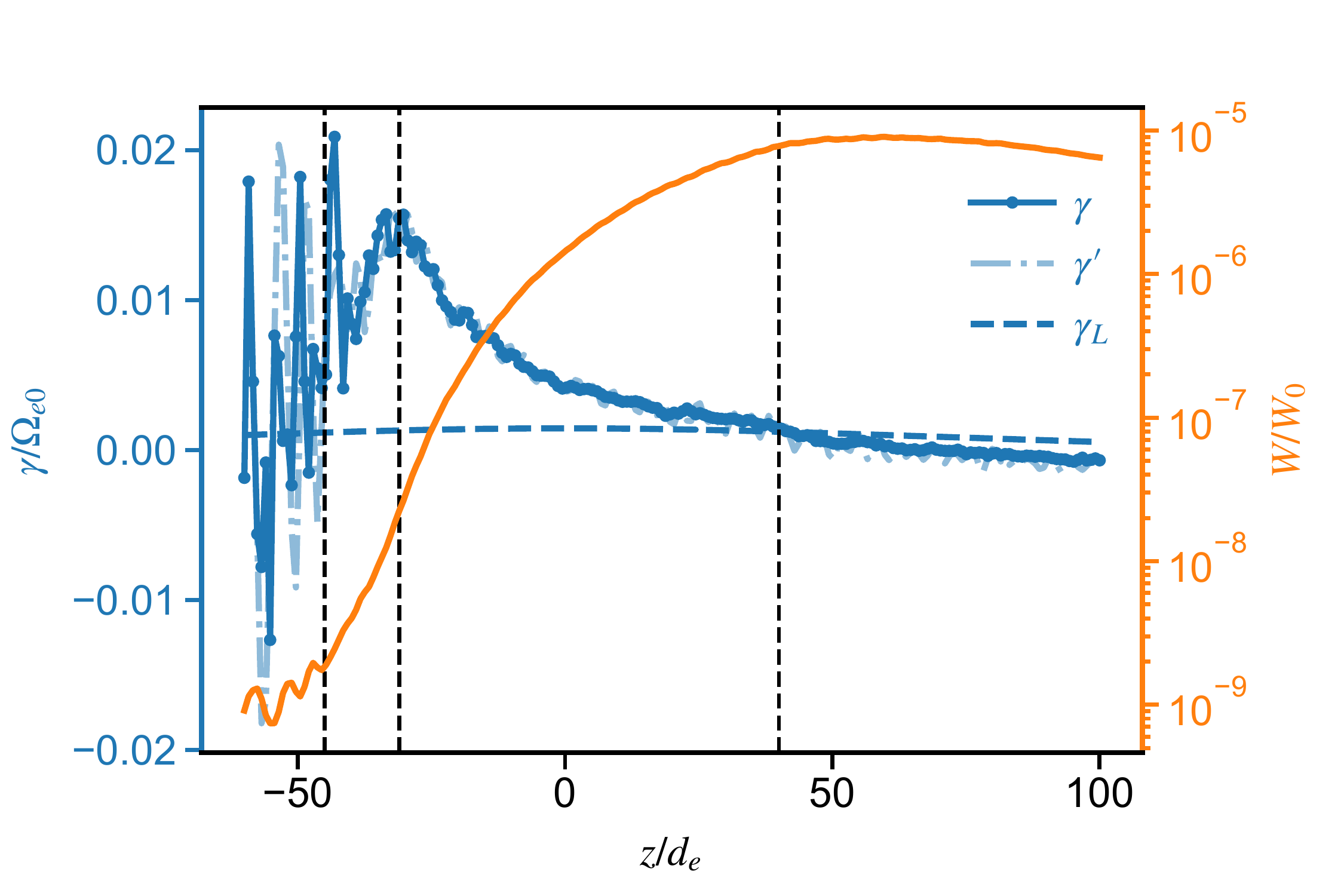}
  \caption{The spatial dependence of effective ($\gamma$ and $\gamma'$) and linear ($\gamma_L$) growth rates, shown in the left y-axis, and the wave energy density ($W$), shown in the right y-axis. The three vertical dashed lines mark the three characteristic locations discussed in the text, from left to right, $z/d_e=-45, -31$, and $40$.}
  \label{fig:gamma_sample}
\end{figure}

To find the ray path, we first locate the time $t_0$ of this ray arriving at $z=0$, which is $t_0\approx 2483 \Omega_{e0}^{-1}$. Then using $t_0$ and $z=0$ as the initial condition, we trace this ray both backward and forward in time to find its ray path for $-60\,d_e \le z \le 100\,d_e$ with 200 $z$'s; i.e., $\Delta z = 0.8\,d_e$. Along its ray path, we calculate $\gamma$ and $\gamma'$ using Equations (\ref{eq:2}) and (\ref{eq:13}), respectively, and the wave energy density $W$ using Equation (\ref{eq:14}). The results are shown in Figure \ref{fig:gamma_sample}. As a reference, we also calculate linear growth rate $\gamma_L$ using the initial bi-Maxwellian distribution. From Figure \ref{fig:gamma_sample}, we note that the wave energy density $W$ fluctuates for $z \lesssim -45\,d_e$, indicated by the left most dashed vertical line, consistent with that noise dominates for $z\lesssim -45\,d_e$ from wave spectrogram (e.g., see Figure \ref{fig:source_identification}). For $z \gtrsim -45\,d_e$, $W$ continuously increases with $z$. Both growth rates $\gamma$ and $\gamma'$ fluctuate at large and negative $z$ and show relatively good agreement with each other for  $z \gtrsim -40\,d_e$. The two growth rates also show a local peak near $z=-31\,d_e$. In our discussion below, we will use $z=-31\,d_e$ as where the peak value of the effective growth rate is for simplicity and consistency. 

Figure \ref{fig:gamma_sample} clearly show three distinctive features of $\gamma$, especially when compared with $\gamma_L$. First, the maximum value of the effective growth rate is in the upstream region near $z=-31\,d_e$, not at the equator where the linear growth rate $\gamma_L$ peaks. Starting from $z=-31\,d_e$, the effective growth rate $\gamma$ decreases as the ray propagates downstream, and becomes smaller than $\gamma_L$ for $z \gtrsim 40\,d_e$. Second, the effective growth rate $\gamma$ is significantly larger than $\gamma_L$ in the upstream, ensuring the growth of a narrowband emission out of linearly unstable broadband whistler mode waves.  Third, the path-integrated convective wave growth rate, characterized by $G \equiv \int \gamma \md z/v_g$, is much larger in the upstream region (from $z=-45\,d_e$ to $z=0$), than that in the downstream region (from $z=0$ to $z=100\,d_e$). This can be directly calculated or can be seen from that $W(z=0)\approx 1.4\times 10^{-6} W_0$, $W(z=-45\,d_e) \approx 1.8\times 10^{-9} W_0$, and $W(z=100\,d_e) \approx 6.3\times 10^{-6} W_0$, where $W_0\equiv B_0^2/8\pi$ is the normalization unit of $W$. Correspondingly, $W(z=0)/W(z=-45\,d_e) \approx 800$ and $W(z=100\,d_e)/W(z=0) \approx 5$. Clearly, the upstream region is significantly more effective at wave amplification than the downstream region. The asymmetry of $\gamma$ and $G$ with respect to $z=0$ suggests that the upstream and downstream interaction regions might play very different roles in chorus wave generation.   

\subsection{Formation of the chirping element using $\gamma$}
\begin{figure}
  \centering
  \includegraphics[width=0.8\textwidth]{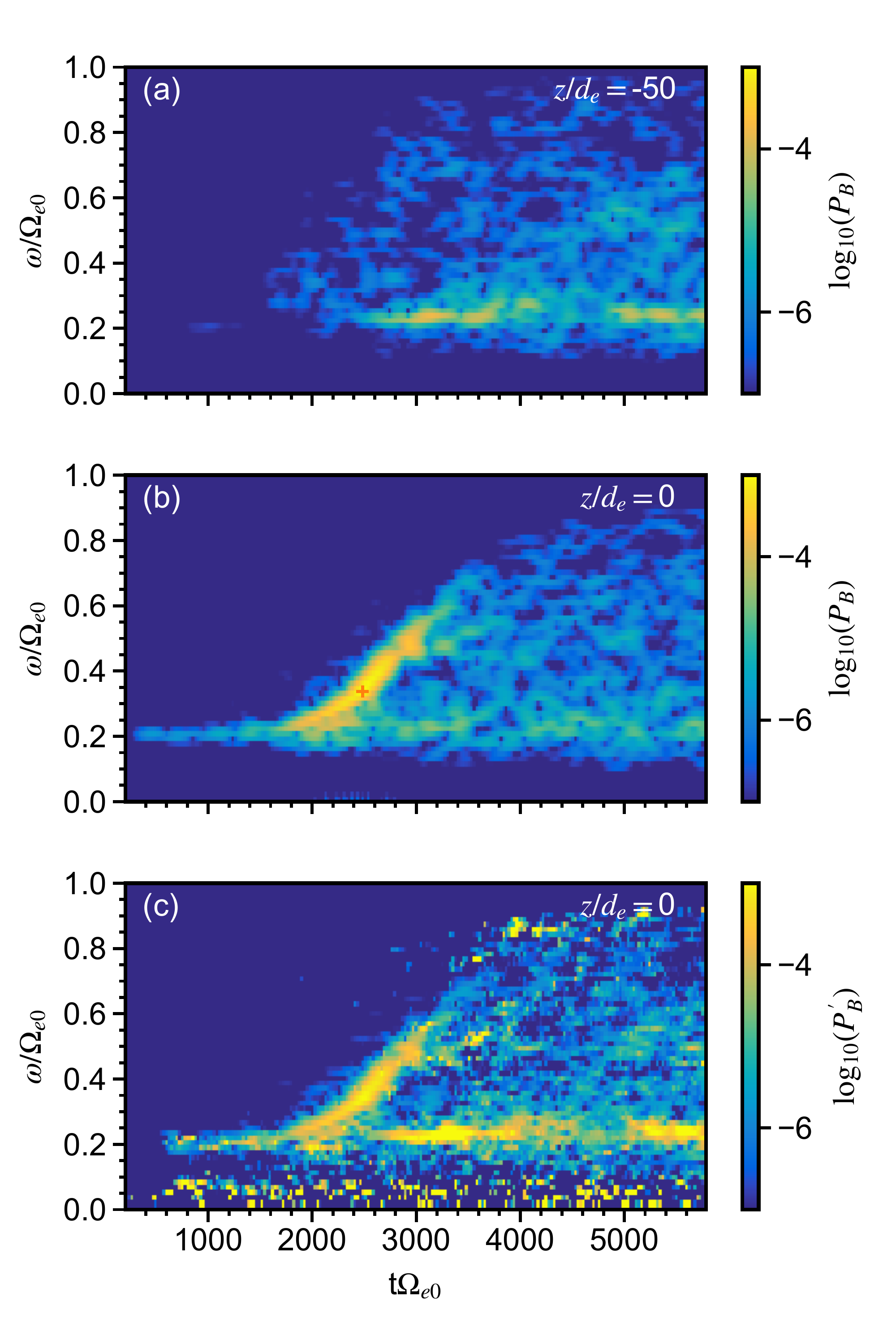}
  \caption{Wave power spectral density ($P_B$) at (a) $z/d_e=-50$ and (b) $0$ from the PIC simulation, and (c) the constructed power spectral density ($P_B'$) using $\gamma$ and $P_B(z/d_e=-50)$. The cross sign in Panel (b) indicates the initial time and frequency used in the calculation of ray path in Figure \ref{fig:gamma_sample}.}
  \label{fig:zonca}
\end{figure}

One of the main principles of the theoretical framework of \citeA{Zonca2017} is that chorus chirping observed at a given location is a result of continuous excitation of narrow spectrum out of a broad and dense background whistler modes. As a consistency check of this principle, we may use the effective growth rate calculated by Equation (\ref{eq:2}) to construct the power spectral density of wave magnetic field at $z=0$ from $P_B(z=-50\,d_e)$ by
\begin{linenomath}
\begin{align}
  \label{eq:4}
  P_B'(\omega, z=0) = P_B(\omega, z=-50\,d_e)\exp\left(2\int_{-50\,d_e}^{0} \gamma/v_g \md z\right).
\end{align}
\end{linenomath}
for all FFT frequencies below $\Omega_{e0}$. It is important to note that, in the above equation, we use $\gamma$ from Equation (\ref{eq:2}) instead of  $\gamma'$ in Equation (\ref{eq:3}). The wave power spectrum at $z=-50\,d_e$ is used because $P_B(z=-50\,d_e)$ related to the chirping element shows basically background thermal noise, as can be seen in Figure \ref{fig:zonca}a. The relatively strong emission near $\omega/\Omega_{e0}\approx 0.2$ at $z=-50\,d_e$ in Figure \ref{fig:zonca}a is due to the weak leftward propagating signal from the antenna being amplified by hot electrons. Applying Equation (\ref{eq:4}) to this signal will produce numerical artifact near $\omega/\Omega_{e0}\approx 0.2$ at $z=0$ because $v_g$ in Equation (\ref{eq:4}) is assumed to be positive in our calculation. This, however, does not affect our conclusions since we focus on reproducing the chorus chirping element, which propagates in $+z$ direction. 

Figure \ref{fig:zonca}b and \ref{fig:zonca}c show the comparison between wave spectrogram from DAWN simulation (Figure \ref{fig:zonca}b) and the spectrum constructed using Equation (\ref{eq:4}) (Figure \ref{fig:zonca}c). Although part of spectrogram is clearly contaminated by noise in the calculated effective growth rate, it is quite obvious that Equation (\ref{eq:4}) is able to reproduce the main portion of the chirping element. This good agreement suggests that our estimated $\gamma$ is quite reliable for most part of the element, but more importantly, it shows consistency with the theoretical framework of \citeA{Zonca2017}. Note that because $v_g$ is independent of time in Equation (\ref{eq:4}), all the dynamics needed to produce frequency chirping is contained in the effective growth rate $\gamma$. Therefore, a plausible scenario is that narrowband emissions of different frequencies are excited at different times and locations, due to the spatial and temporal dependence of $\gamma$, in the upstream region. These narrowband emissions arrive at a given location at downstream at different times, leading to a chirping element with increasing frequency. A self-consistent description of how the nonlinear wave particle interactions modify the hot electron distribution function and the effective growth rate is given in the theoretical framework of \citeA{Zonca2017} and \citeA{Zonca2021}. 

\subsection{Identification of the source region}
\label{sec:ident-source-regi}
The spatial profile of $\gamma$ and $W$ along the ray path shown in Figure \ref{fig:gamma_sample} suggests that the generation of emission with $\omega=0.34\Omega_{e0}$ occurs somewhere near $z/d_e=-40\sim -50$. This could be further confirmed by plotting the spectrogram of waves at various locations. The principle of using wave spectrogram to constrain the source location is simple: for a given $z_0$, if a clear signal of the given frequency $\omega=0.34\Omega_{e0}$ along the ray path is identifiable, then this particular mode is generated at a location $z< z_0$, since waves of interest propagate in $+z$ direction. 

\begin{figure}
  \centering
  \includegraphics[width=1.0\textwidth]{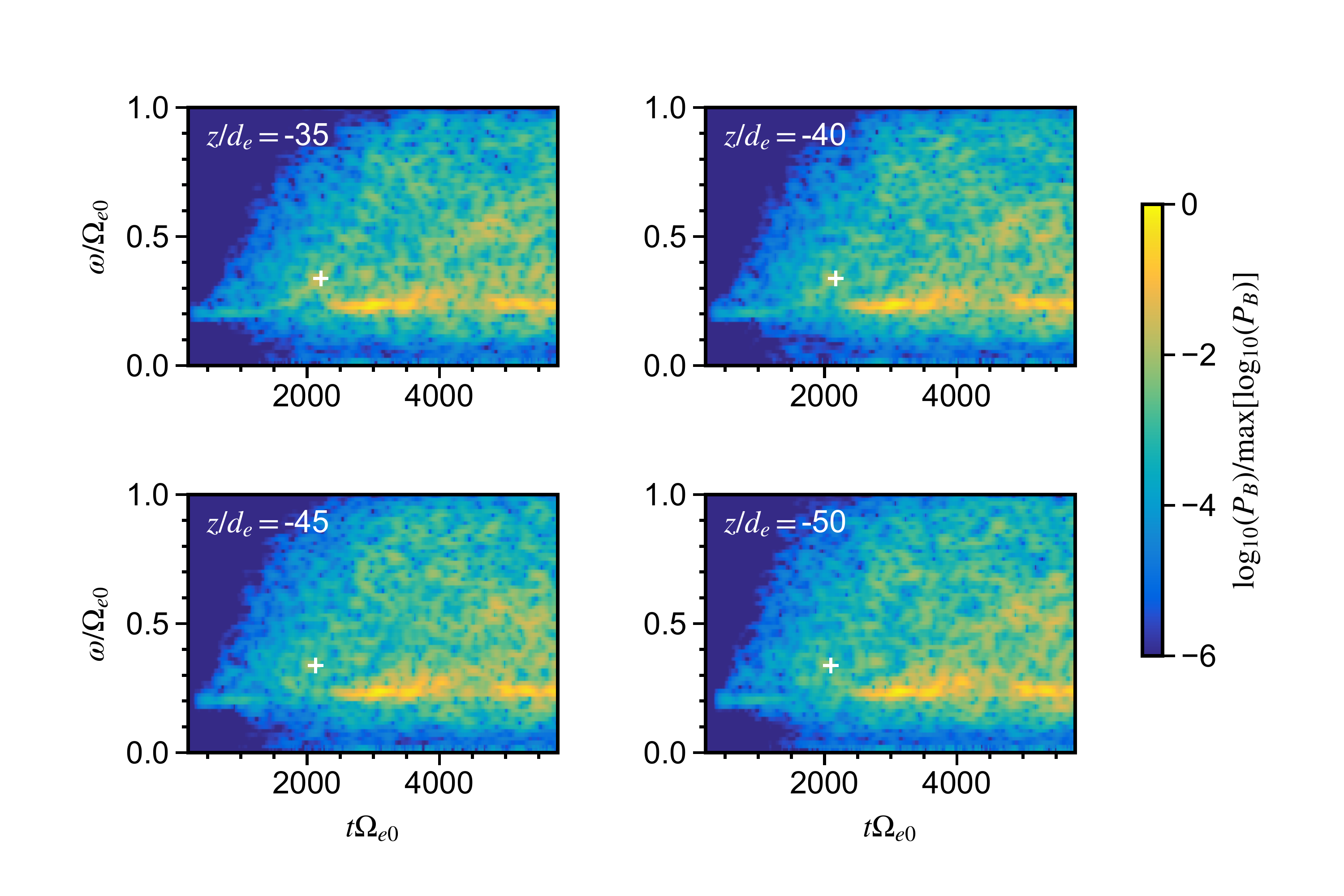}
  \caption{Wave power spectrograms at four different locations ($z/d_e$): (a) $-35$, (b) $-40$, (c) $-45$, (d) $-50$. These spectrograms are used to help identify of the source region of the emission with $\omega/\Omega_{e0}=0.34$. The white cross marks the time and location of the ray at these four locations.}
  \label{fig:source_identification}
\end{figure}

Figure \ref{fig:source_identification} shows four spectrograms at $z/d_e=-35, -40, -45$, and $-50$. In Figure \ref{fig:source_identification}a, $z/d_e=-35$, we see a clear part of the chirping element including $\omega=0.34\Omega_{e0}$.  This element is also weakly visible at $z/d_e=-40$ (Figure \ref{fig:source_identification}b). These two spectrograms suggest that the source location of the narrowband emission with $\omega=0.34\Omega_{e0}$ is at $z/d_e < -40$. It is difficult to identify a chirping element at $z/d_e=-45$ in Figure \ref{fig:source_identification}c, and it is safe to say that at $z/d_e=-50$ (Figure \ref{fig:source_identification}d), the spectrogram along the ray path of the element shows basically thermal noise.  

Combining the spectrograms in Figure \ref{fig:source_identification} and the variation of $\gamma$ and $W$ shown in Figure \ref{fig:gamma_sample}, we conclude that the source region for generation of the emission with $\omega=0.34\Omega_{e0}$ is located roughly between $z/d_e= -40$ and $-50$.   

\subsection{Variation of $R$ for resonant electrons }
Having identified the source region, we now analyze the $R$ parameter at wave generation. This parameter is important in studying nonlinear resonant wave particle interactions and is defined through the second order time derivative of wave particle interaction phase angle $\zeta \equiv \langle \vec{v}_\perp, \delta \B \rangle$. For resonant interactions between electrons and parallel whistler waves, the first order time derivative of $\zeta$ is
\begin{linenomath}
\begin{align}
  \frac{\mathrm{d}\zeta}{\mathrm{d} t} = \omega-k v_\| -\Omega_e
\end{align}
\end{linenomath}
and its second order time derivative is
\begin{linenomath}
\begin{align}
  \label{eq:15}
  \frac{\mathrm{d}^2\zeta}{\mathrm{d} t^2} &= \omega_{tr}^2 \sin\zeta - (R_1+R_2),
\end{align}
\end{linenomath}
where
\begin{linenomath}
\begin{align}
  \label{eq:6}
  R_1 &= \left(1 - \frac{v_r}{v_g}\right)^2 \frac{\partial \omega}{\partial t}, \\
  \label{eq:7}
  R_2 &=\left(\frac{k v_{\perp}^2}{2\Omega_e} - \frac{3v_r}{2}\right) \frac{\partial \Omega_e}{\partial z}.
\end{align}
\end{linenomath}
Clearly $R_1$ characterizes effects of frequency chirping; $R_2$, effects of background magnetic field inhomogeneity. In above equations, $v_r$ is the resonant parallel velocity that gives $\mathrm{d} \zeta/\mathrm{d} t =0$, $v_\perp$ is the perpendicular velocity with respect to background $\B$, and $\omega_{tr}^2 = k v_\perp e \delta B/ mc$ is the characteristic phase-trapping frequency. For convenience, we may also write $\omega_{tr}$ using normalized variables as
\begin{linenomath}
\begin{align}
  \label{eq:1}
  \left(\frac{\omega_{tr}}{\Omega_e}\right)^2 = \mu \left(\frac{\omega}{\Omega_e}\right) \left( \frac{v_\perp}{c} \right) \left(\frac{\delta B}{B}\right).
\end{align}
\end{linenomath}
where $\mu\equiv ck/\omega$ is the refractive index. Relativistic effects could easily be included if needed. Equation (\ref{eq:1}) should be handy to quickly estimate $\omega_{tr}$ in various cases. The parameter $R$ is defined by $R = (R_1+R_2)/\omega_{tr}^2$.

\begin{figure}
  \centering
  \includegraphics[width=1.0\textwidth]{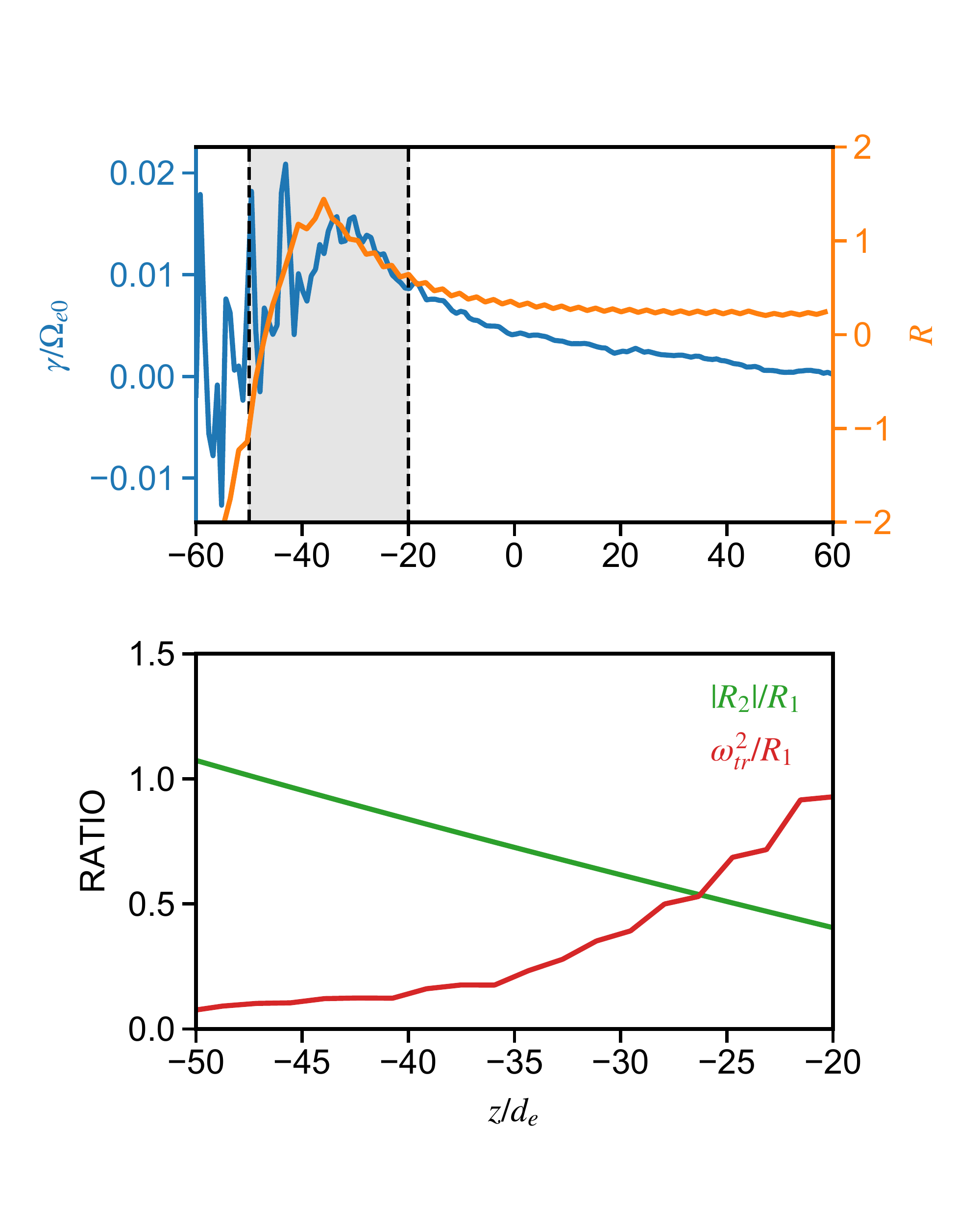}
  \caption{(Top) The spatial dependence of effective growth rate $\gamma$ (the left y-axis) and $R$ (the right y-axis) for $-60 \leq z/d_e \leq 60$. (Bottom) The spatial variation of the ratio $|R_2|/R_1$ (green) and $\omega_{tr}^2/R_1$ (red) for $-50 \leq z/d_ \leq -20$, the shaded region marked by the vertical dashed lines in the top panel.}
  \label{fig:R}
\end{figure}

The top panel of Figure \ref{fig:R} shows the variation of $R$ together with the effective growth rate for $| z/d_e| \leq 60$, and the shaded region represents the range of $z$ ($-50 \leq z/d_e \leq -20$) where ratios of $|R_2|$ and $\omega_{tr}^2$ to $R_1$ are shown in the bottom panel. In calculation of $R$, we have used $\partial \omega/\partial t = 2.6\times 10^{-4} \Omega_{e0}^{-2}$ estimated from wave spectrogram, and $\alpha \equiv \tan^{-1}(v_\perp/v_\|)= 70\degree$, since it is most easily for phase-trapping to occur for particles with pitch angles near $65-75$\degree \cite{Inan1978}. 

Figure \ref{fig:R} shows a few interesting features of $R$ and the underlying electron dynamics. First, from $z=0$ to $60\,d_e$, the value of $R$ keeps decreasing from $0.35$ to about $0.2$, suggesting stronger nonlinear wave particle interactions in the downstream than at the equator. Second, between $z=-30\,d_e$ and about $-40\,d_e$, $R \gtrsim 1$, suggesting that the wave is not effective at phase-trapping resonant electrons because of its weak amplitude. The value of $R$ keeps decreasing to $0$ and negative values as $z$ further decreases. However, this does not mean that phase-trapping is possible again, because waves at these locations are broadband (e.g., see Figure \ref{fig:source_identification}d). The reason for $R$ becoming $0$ or negative can be seen from the bottom panel of Figure \ref{fig:R}, which shows $|R_2|/R_1$ and $\omega_{tr}^2/R_1$. From $z/d_e=-20$ to $-50$, $\omega_{tr}^2/R_1$ decreases and $|R_2|/R_1$ increases, because of decreasing wave amplitude and increasing background field inhomogeneity, and the two ratios become comparable near $z/d_e\approx -26$. From $z/d_e= -40$ to $-50$, which is about the source region of the emission with $\omega=0.34$, $|R_2|/R_1$ changes from about $0.8$ to $1$ and $\omega_{tr}^2/R_1$ changes from about  $0.1$ to $0.05$. Therefore, the value of $R$ being close to $0$ near $z/d_e\sim -50$ simply because $R_1$ nearly cancels $R_2$. From $z/d_e=-40$ to $-50$,  we clearly have the following ordering between $R_1$, $R_2$ and $\omega_{tr}^2$: $R_1/R_2 \sim -1$ and $R_1, R_2 \gg \omega_{tr}^2$. This is fully consistent with results of \citeA{Wu2020b}, and suggests that
\begin{linenomath}
\begin{align}
  \label{eq:11}
  \frac{\mathrm{d}^2\zeta}{\mathrm{d} t^2} \approx - (R_1+R_2) \approx 0,
\end{align}
\end{linenomath}
from Equation (\ref{eq:15}). Physically, this means that the phase-locking condition ($\mathrm{d}^2\zeta/\mathrm{d} t^2=0$) could be satisfied through balancing $R_2$ by $R_1$. As discussed below in Section \ref{sec:trap-release-tar}, this can also be understood as a result of selective amplification of new emissions that can stay resonant ($\mathrm{d}\zeta/\mathrm{d} t=0$) with these electrons for the longest possible time because of phase locking.

\section{Electron phase space dynamics}
\label{sec:electron-phase-space}
\subsection{Entrapping and detrapping of resonant electrons}
\label{sec:entr-detr-reson}
Previous section analyzes the effective growth rate $\gamma$ along a given ray path ($\omega=0.34\Omega_{e0}$). Here we focus on a group of selected electrons, and demonstrate their trapping and release from the chorus wave packet in the downstream and upstream, respectively. We save $55$ electron distributions equidistantly in time from $t\Omega_{e0}=1500$ to $5500$, or one distribution every $\sim 74\Omega_{e0}^{-1}$. One of the saved electron distributions is at $t\Omega_{e0}= 2241$, when $\gamma$ of $\omega=0.34\Omega_{e0}$ peaks near $z/d_e=-31$. 

\begin{figure}[!h]
  \centering
  \includegraphics[width=\textwidth]{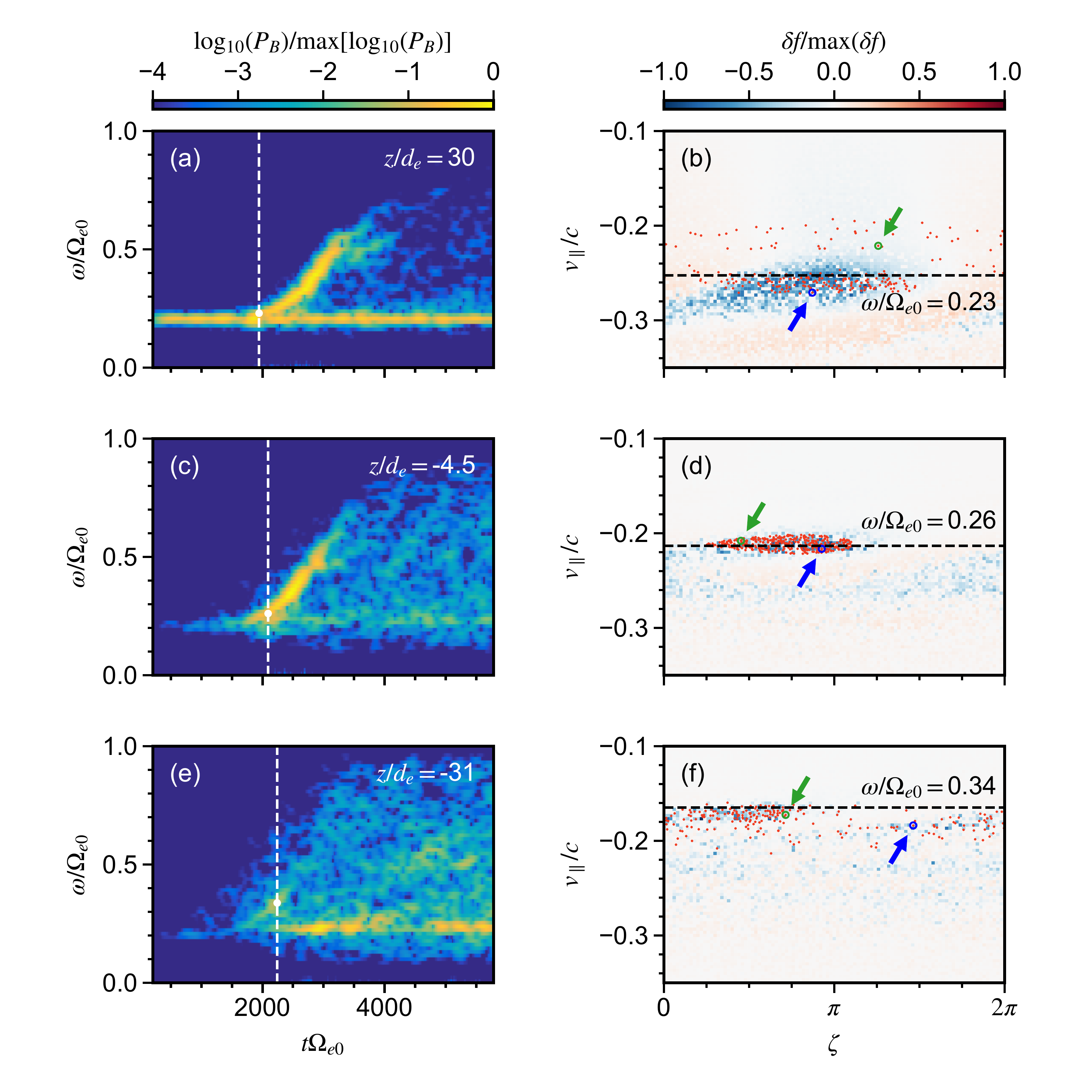}
  \caption{(Left) Wave spectrograms at three different locations ($z/d_e$): (a) $30$, (c) $-4.5$, and (e) $-31$. (Right) The corresponding electron phase space ($v_\|$-$\zeta$) distributions. In the left column, the vertical dashed lines mark the time of the corresponding phase space distribution. In the right column, the horizontal dashed lines mark the frequency of maximum wave intensity, and red dots mark phase space coordinates of selected electrons. The blue and green circles and arrows mark two electrons whose energy and phase angle variation are highlighted in Figure \ref{fig:psi_change}.}
  \label{fig:electron_dynamics}
\end{figure}

Figure \ref{fig:electron_dynamics} illustrates both the wave spectrograms (left column) and the phase space distributions of a group of selected electrons ($375$ in total) represented by red dots (right column) at three different locations. The criterion for selecting electrons is that they are located within the resonant island at $z/d_e=-4.5$ and $t\Omega_{e0}=2092.6$ (Figure \ref{fig:electron_dynamics}d). This particular time and location are used because most of these electrons would reach $z/d_e=-31$ near $t\Omega_{e0}=2241$, responsible for the peaking of the growth rate of the mode with $\omega=0.34\Omega_{e0}$. Backward from $z/d_e=-4.5$ by an equal amount of time, most electrons were located at $z/d_e = 30$, as shown in Figure \ref{fig:electron_dynamics}b. The spectrograms in the left column show the frequency of maximum intensity of the chirping element at the three different times, indicated by vertical dashed lines. From the phase space plots of resonant islands in the right column, these frequencies are also resonant frequencies. 

Figure \ref{fig:electron_dynamics}b, \ref{fig:electron_dynamics}d, and \ref{fig:electron_dynamics}f clearly illustrate how electrons get phase-trapped and released as they move from downstream to upstream and resonantly interact with a chorus packet with chirping frequency. At $z/d_e=30$, only part of the selected electrons are phase-trapped and located within the resonant island of the wave with $\omega=0.23\Omega_{e0}$. 
Because resonant electrons move in opposite direction to the chirping wave packet, as the selected electrons move upstream, the resonant frequency increases. At $z/d_e=-4.5$, all electrons are located within the resonant island and phase-trapped by the wave with $\omega=0.26\Omega_{e0}$.  Clearly, from $z/d_e=30$ to $-4.5$, the selected electrons experienced an entrapping process and become phase-organized. As the electrons arrive at $z/d_e=-31$ and the resonant frequency becomes $\omega/\Omega_{e0}=0.34$, however, they start to get de-trapped as the calculated $R\sim 1$, although the selected electrons are still phase-correlated (Figure \ref{fig:electron_dynamics}f). It is also clear that these selected electrons amplify the emission with $\omega/\Omega_{e0}=0.34$, but are not responsible for its generation (see Section \ref{sec:ident-source-regi}). As they move further upstream, the electrons will resonantly interact with waves whose frequency is higher than  $0.34\Omega_{e0}$. Finally, we would likely to point out that, although the dynamics of only a selected group of electrons are analyzed, there is a continuous entrapping of fresh electrons in the downstream and release of these electrons in the upstream, responsible for continuous generation of new emissions. This mechanism can be viewed as the chorus wave packet slipping over the population of resonant electrons maximizing wave particle power extraction, and calls for the analogy with super-radiance in free electron lasers introduced by \citeA{Zonca2015} discussing energetic particle mode convective amplification in fusion plasmas. We shall come back to this point in the next section.

\subsection{Variation of wave particle interaction phase angle}
The dynamics of resonant electrons demonstrated above clearly show similarities between chorus generation and various electronic devices, such as backward wave oscillators and free electron lasers. In fact, several authors have built chorus wave models based on the understanding of these devices which are capable of generating coherent emissions \cite{Trakhtengerts1995,Soto-Chavez2012}. In studies of wave generation in these devices, one important variable is how the wave particle interaction phase angle ($\zeta$) changes during the whole interaction process. Here we show the variation of $\zeta$ of the selected electrons as they move from $z/d_e = 30$ to $-31$, along with the energy change, as an indicator of phase-trapping. For simplicity of discussion, two electrons are selected, as indicated by the green and blue circles and arrows in Figure \ref{fig:electron_dynamics}. These two electrons are from slightly different groups of selected electrons, and will be called blue and green electrons, respectively. The blue electron represents those that are already phase-trapped at $z/d_e=30$, and the green electron represents those that are untrapped at $z/d_e=30$ but get phase-trapped somewhere between $z/d_e=30$ and $-4.5$. 

\begin{figure}
  \centering
  \includegraphics[width=\textwidth]{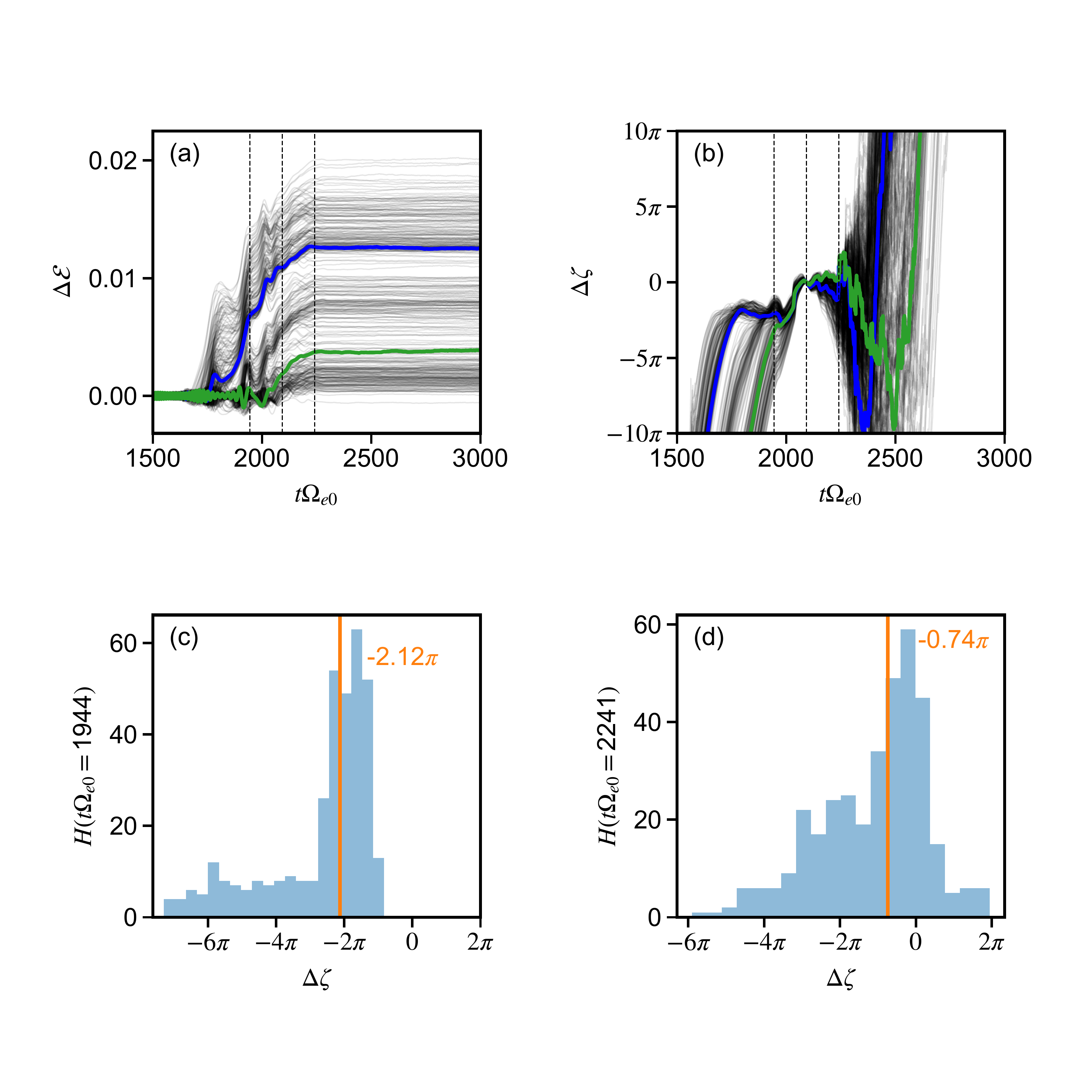}
  \caption{\label{fig:psi_change} Variation of energy ($\mathcal{E}$) and $\zeta$ of all selected electrons. (a) Energy change from its initial value.  (b) Phase angle change with respect to $\zeta(z/d_e=-4.5)$. (c) Histograms of $\Delta \zeta$ at $z/d_e=30$.  (d) Similar to (c), but at $z/d_e=-31$. In (a) and (b), the blue and green lines are for the electrons marked by blue and green circles in Figure \ref{fig:electron_dynamics}.}
\end{figure}

Figure \ref{fig:psi_change}a and \ref{fig:psi_change}b show the energy change, $\Delta \mathcal{E} \equiv \mathcal{E}(t)- \mathcal{E}(t=0)$, and the $\zeta$ change, $\Delta \zeta\equiv \zeta(z)-\zeta(z=-4.5)$, of all selected electrons, respectively. Note that different types of reference values are used for $\Delta \mathcal{E}$ and $\Delta \zeta$: $\Delta \mathcal{E}$ is the energy change from its initial value, while $\Delta \zeta$ is the phase angle change with respect to $z/d_e=-4.5$, near the equator. The energy change of the blue electron in Figure \ref{fig:psi_change}a suggests that this electron got phase-trapped at about $t\Omega_{e0}\approx 1700$. Similarly,  the green electron got phase-trapped at about $t\Omega_{e0}\approx 2000$. This is consistent with the much slower change of $\zeta$ for the blue (green) electron from $t\Omega_{e0}\approx 1700$ ($2000$) to about $2241$ than the rest of the time. Figure \ref{fig:psi_change}a also suggests that $\Delta \mathcal{E}$ is much larger for the blue electron than for the green electron, due to longer time of phase-trapping. One thing worth pointing out here is the much larger energy oscillation near $t\Omega_{e0}\approx 1700$ than at later times. This is consistent with the much larger amplitude modulation of waves in the downstream region as discussed in Section \ref{sec:fine-struct-chor}.

Figure \ref{fig:psi_change}c and \ref{fig:psi_change}d show the histogram of $\Delta \zeta$ at $z/d_e = 30$ and $-31$. The median value of $\Delta \zeta$ is $-2.12\pi$ at $z/d_e=30$ and $-0.74\pi$ at $z/d_e=-31$. The large variance of both statistics is associated with the fact that only part of the electrons are phase-trapped at $z/d_e=30$ and part of the electrons get de-trapped slightly before reaching $z/d_e=-31$. Note the definition of $\Delta \zeta$, we may conclude that as electrons move from $z/d_e=30$ to $-4.5$, $\zeta$ increases by about $2\pi$, and as they move from $z/d_e=-4.5$ to $-30$, $\zeta$ decreases by about $0.7\pi$.

These results are consistent with the analysis presented in Section \ref{sec:entr-detr-reson} and can be interpreted as electrons in the downstream region forming a phase space structure when moving from $z/d_e = 30$ to $z/d_e = -4.5$ (cf. Figure \ref{fig:electron_dynamics}) as they slip over the propagating chorus wave packet. Then, they are released from the structure as they reach $z/d_e = -31$ being phase shifted by $\sim -\pi$. The close analogy with super-radiance in free electron lasers \cite{Zonca2015,Zonca2015a} is apparent when comparing panels (b) and (d) of Figure \ref{fig:electron_dynamics} with panels (a) and (b) of Figure 3 in \citeA{Giannessi2005}. Note that, in the case of Figure \ref{fig:psi_change}, we have focused the attention on two particles gaining energy from the interaction with the chorus element. However, the majority of electrons in the downstream region do the opposite, resulting in the overall amplification of the wave-packet; similar, again, to the free electron laser case.

\section{A ``Trap-Release-Amplify'' (TaRA) model}
\label{sec:trap-release-tar}
Combing results from Sections \ref{sec:effect-growth-rate} and \ref{sec:electron-phase-space}, we now describe a phenomenological model of chorus, called the ``Trap-Release-Amplify'' (TaRA) model. This model explains how chirping occurs and relates the chirping rate of chorus to both the background magnetic field inhomogeneity and to wave amplitude. 

\begin{figure}
  \centering
  \includegraphics[width=0.9\textwidth]{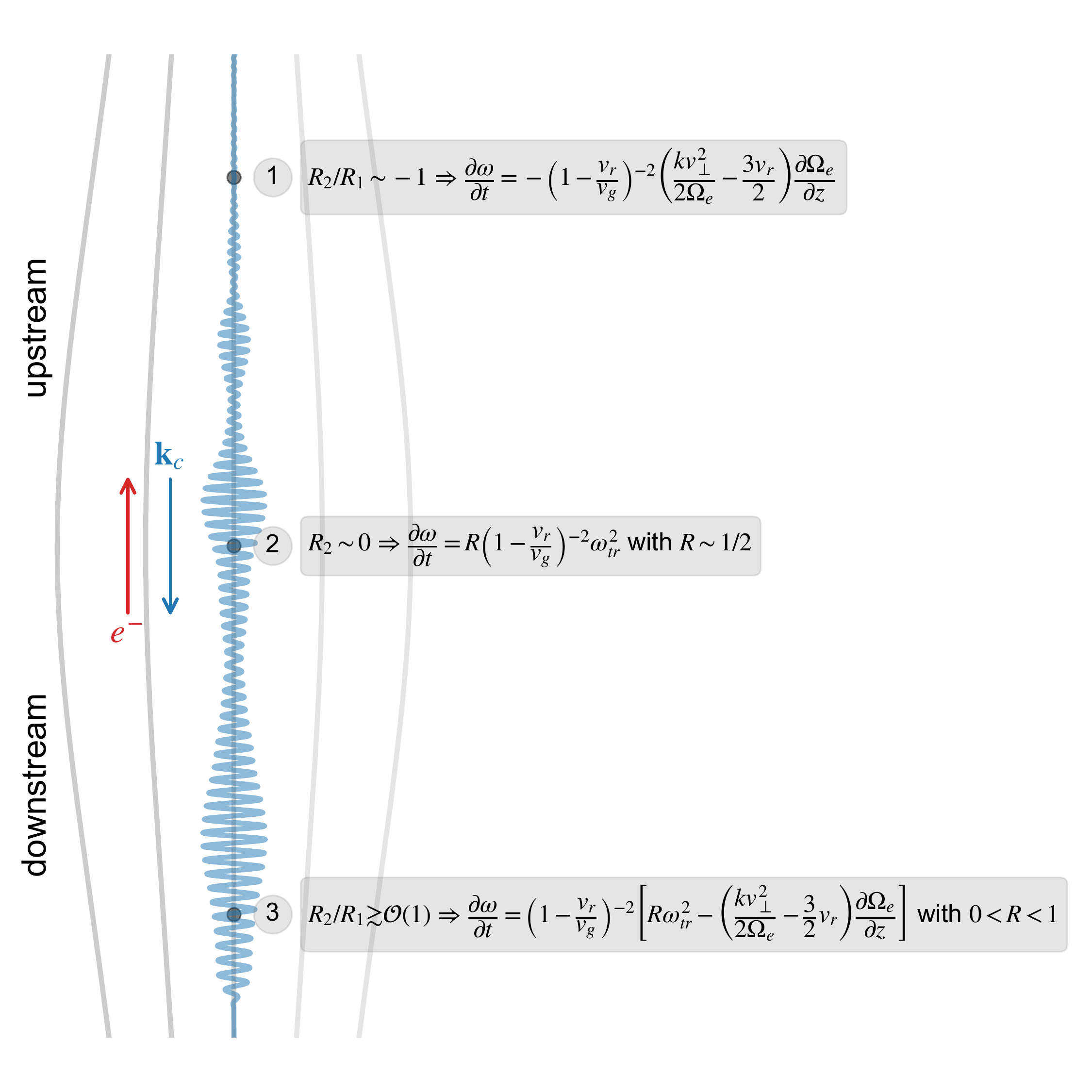}
  \caption{Illustration of the TaRA model. The red arrow indicates the motion of resonant electrons ($e^-$), while the blue arrow indicates the direction of wave vector ($\vec{k}_c$) of chorus. The blue waveform is taken from simulation. Points 1, 2, 3 represent the release point ($R_2\gg \omega_{tr}^2$), the equator ($R_2=0$), and a point in the downstream where $R_2/R_1 \gtrsim \mathcal{O}(1)$. The corresponding equations for the chirping rate are also given at the three points.}
  \label{fig:tar}
\end{figure}

Figure \ref{fig:tar} illustrates the model, which shows the waveform of chorus at a given time taken from simulation, along with the background magnetic field. We assume that a chorus wave packet consists of a nearly continuous spectrum of whistler modes $\omega_0, \omega_1, \omega_2, \cdots $, with $\omega_0 < \omega_1 < \omega_2 < \cdots$. The spacing between these frequencies is infinitely small. We further assume that the generated part of the packet consists of modes with frequency from $\omega_0$ to $\omega_N$, and modes with $\omega_{i}, i > N$ are yet to be exited. For a rising-tone chirping element, clearly $\omega_0$ is located in the front of the packet; $\omega_N$, the rear of the packet.

Because of the opposite movement of resonant electrons and wave packet, the resonant interaction between fresh electrons and the generated part of the chorus packet starts from $\omega_0$ in the downstream. These electrons get phase-trapped in the interaction process as they move toward the upstream, producing a ``phase-bunched'' current. As the electrons move to a region in the upstream such that the wave amplitude is too small to continue phase-trapping, the resonant electrons are released. Note that these electrons are still phase-correlated, and once released, they can selectively amplify new emissions ($\omega > \omega_N$) from the broadband background whistler wave spectrum based on the phase-locking condition: $\mathrm{d}^2\zeta/\mathrm{d} t^2 = 0$. This selection rule ensures that the resonance condition $\mathrm{d}\zeta/\mathrm{d} t = 0$ can be maintained for the longest possible time, therefore maximizes power transfer to the wave. 

Based on this model, a few equations about chorus frequency chirping rate can be obtained. First, at the release point (Point 1 in Figure \ref{fig:tar}) where new emissions are excited, the wave amplitude term $\omega_{tr}^2$ is much smaller than the nonuniformity term, $R_2$. Correspondingly, the phase locking condition requires balancing $-R_2$ by $R_1$; i.e.,
\begin{linenomath}
\begin{align}
  \frac{\mathrm{d}^2\zeta}{\mathrm{d} t^2} = 0 \Rightarrow R_1 \sim -R_2, \text{ if } R_2 \gg \omega_{tr}^2, 
\end{align}
\end{linenomath}
which leads to,
\begin{linenomath}
\begin{align}
  \label{eq:5}
  \pd{\omega}{t} = -\left(1 - \frac{v_r}{v_g}\right)^{-2} \left(\frac{k v_{\perp}^2}{2\Omega_e} - \frac{3v_r}{2}\right) \frac{\partial \Omega_e}{\partial z},
\end{align}
\end{linenomath}
where we have used definitions of $R_1$ and $R_2$ in Equations (\ref{eq:6})-(\ref{eq:7}). Equation (\ref{eq:5}) clearly defines a relation between the chirping rate and the background magnetic field inhomogeneity. In fact, according to our discussion above, we may say that, for the particular case of the rising-tone emission with $k>0$, the negative inhomogeneity in the upstream region is the reason for the upward chirping of frequency. Second, the nonlinear phase-trapping of electrons caused by the generated part of the packet mainly occurs in the downstream region. Near the equator, $R_2\sim 0$ (Point 2 in Figure \ref{fig:tar}). Based on the study of \citeA{Vomvoridis1982}, effective wave power transfer typically occurs for $R$ between about $0.2$ and $0.8$. For simplicity, $R$ is taken to be $1/2$ as in \citeA{Vomvoridis1982}. Similar conclusion has also been reached by \citeA{Omura2008} ($R=-0.4$), based on an assumed form of distribution for phase trapped electrons, and \citeA{Zonca2017}, based on a self-consistent theoretical framework of chorus. This leads to the well-known relation between the frequency chirping rate and the wave amplitude,
\begin{linenomath}
\begin{align}
  \label{eq:12}
  \pd{\omega}{t} = R \left(1-\frac{v_r}{v_g}\right)^{-2} \omega_{tr}^2, \text{ with } R\sim \frac{1}{2}.
\end{align}
\end{linenomath}
We must emphasize that, however, new emissions of the chorus packet are generated at Point 1 in Figure \ref{fig:tar}, not at the equator. Correspondingly, Equation (\ref{eq:5}) describes how frequency chirping occurs and why it is rising-tone in the current case, while Equation (\ref{eq:12}) emphasizes the importance of nonlinear wave particle interactions in chorus generation. Finally, in the downstream region at Point 3, where $R_2$ is comparable to or larger than $R_1$, the contribution of the inhomogeneity term in calculation of $R$ cannot be ignored, resulting in that
\begin{linenomath}
\begin{align}
  \pd{\omega}{t} = \left(1-\frac{v_r}{v_g}\right)^{-2} \left[R \omega_{tr}^2-\left(\frac{k v_{\perp}^2}{2\Omega_e} - \frac{3v_r}{2}\right) \frac{\partial \Omega_e}{\partial z}\right],
\end{align}
\end{linenomath}
with $0<R<1$ if the wave particle interaction is nonlinear. 

From our description above, it is clear that the TaRA model shares a basic principle with most of previously published models of chorus \cite{Helliwell1967, Nunn1974,Vomvoridis1982,Omura2008}; i.e., nonlinear wave particle interactions phase organize resonant electrons, allowing generation of narrowband coherent emissions. The difference among various models is mainly about how this principle is applied to explain chirping, to determine the chirping rate, and to explain various fine structures of chorus. Before detailed comparisons with previous models in Section \ref{sec:comp-with-prev}, we would like to point out that, in the TaRA model, the chirping rate is related to both the background magnetic field inhomogeneity as by \citeA{Helliwell1967} in the upstream source region and to the wave amplitude as by \citeA{Vomvoridis1982} at the equator. Correspondingly, our model unifies previous conclusions from two seemingly unrelated models \cite{Helliwell1967,Vomvoridis1982} for the first time, as far as we are aware of. The difference in two ways of estimating chirping rate, Equations (\ref{eq:5}) and (\ref{eq:12}), is mainly caused by that they are derived at different stages of chorus generation, as illustrated in Figure \ref{fig:tar}. 

\section{Comparison with previous models}
\label{sec:comp-with-prev}
In this section, we compare in detail the TaRA model with two previous models. One is Helliwell's model \cite{Helliwell1967}, because the ``consistent-wave'' condition assumed by Helliwell's model to explain chirping is simply a special case of the general phase-locking condition used by the TaRA model, despite differences in other aspects of the models. The second one is the sequential triggering model \cite{Omura2011}, because it is one of the most influential models in recent years. 

\subsection{Helliwell's model}
 \label{sec:helliwells-model}
 First, we demonstrate that Equation (\ref{eq:5}) could be written in exactly the same form as the one given by \citeA{Helliwell1967}. Using the resonance condition $\omega-kv_r = \Omega_e$, we can write
\begin{linenomath}
\begin{align}
  \label{eq:9}
  -\left(\frac{kv_\perp^2}{2\Omega_e}-\frac{3}{2} v_r\right) = \frac{3v_r}{2} \left(1+\frac{1-\omega/\Omega_e}{3}\tan^2\alpha \right).
\end{align}
\end{linenomath}
With $v_g/v_r = -2\omega/\Omega_e$ for $\mu \gg 1$, it is straightforward to demonstrate that
\begin{linenomath}
\begin{align}
  \label{eq:8}
  \left(1 - \frac{v_r}{v_g}\right)^{-2} = -\frac{3\omega/\Omega_e}{1+2\omega/\Omega_e} \frac{v_g}{1-v_g/v_r}\frac{2}{3v_r}.
\end{align}
\end{linenomath}
Substituting Equations (\ref{eq:9}) and (\ref{eq:8}) into Equation (\ref{eq:5}), we arrive at
\begin{linenomath}
\begin{align}
  \label{eq:10}
  \pd{\omega}{t} = -\frac{v_g}{1-v_g/v_r}\frac{3\omega/\Omega_e}{1+2\omega/\Omega_e} \left(1+\frac{1-\omega/\Omega_e}{3}\tan^2\alpha \right) \frac{\partial \Omega_e}{\partial z}.
\end{align}
\end{linenomath}
This equation is in exactly the same form as Equation (14) of \citeA{Helliwell1967}, after taking into consideration the difference about signs of $v_g$ and $v_r$ in this paper and \citeA{Helliwell1967}.

The agreement between Equation (\ref{eq:5}) and (\ref{eq:10}) is not surprising if one realizes that the phase locking condition with $R_2\gg \omega_{tr}^2$ is exactly the same as the ``consistent-wave'' condition used by \citeA{Helliwell1967}, who requires that the Doppler shifted wave frequency changes to match the spatial variation of the background electron gyrofrequency. The two models, therefore, share the same principle that frequency chirping results from the maximization of wave power transfer. On the other hand, Helliwell's model could be regarded as a special case of the TaRA model when nonuniformity ($R_2$) dominates nonlinearity ($\omega_{tr}^2$). This could be seen from that Helliwell's model does not allow chirping when the background magnetic field is uniform ($R_2=0$), as can be seen from Equation (\ref{eq:10}). However, in this case, it is still possible for $R_1$ to be balanced by the $\omega_{tr}^2$ term to achieve phase-locking in the TaRA model. The phase locking condition ($\mathrm{d}^2\zeta/\mathrm{d} t^2=0$) cannot be reduced to Equation (\ref{eq:11}), but is equivalent to
\begin{linenomath}
\begin{align}
  \frac{\mathrm{d}^2\zeta}{\mathrm{d} t^2} &= \omega_{tr}^2 \sin\zeta - R_1 = 0.
\end{align}
\end{linenomath}
This suggests the possibility of frequency chirping in a uniform background magnetic field, as demonstrated by \citeA{Wu2020b} and discussed by \citeA{Zonca2021}. Correspondingly, the chirping in a uniform magnetic field might be considered as the case where nonlinearity ($\omega_{tr}^2$) dominates nonuniformity ($R_2$), opposite to the case discussed by \citeA{Helliwell1967} and the rest part of this paper. Application of the TaRA model to chirping in a uniform field is left as a future study. 

In general, the location $z_0$, used to evaluate $\partial\Omega_e/\partial z$ in Equations (\ref{eq:10}) or (\ref{eq:5}), needs to be calculated by taking into account the whole history of nonlinear wave particle interactions in both downstream and upstream. However, because electrons mainly get de-trapped in the upstream and the wave field in the upstream is weak, it is possible to assume adiabatic motion of electrons in the upstream and obtain the lowest order estimate of $z_0$, as done by \citeA{Helliwell1967}. Correspondingly, we may use $|z_0| = (2\pi |v_\||/\xi \Omega_{e0})^{1/3}$ from \citeA{Helliwell1967} in Equation (\ref{eq:5}) or Equation (\ref{eq:10}). This value of $z_0$ corresponds to a change of the wave particle interaction phase angle by $\pi$ due to adiabatic motion of electrons and the background magnetic field nonuniformity (see also Figure \ref{fig:psi_change}). For the case in this study, $z_0 \approx -39\,d_e$, given that $v_r/c \approx -0.2$ for  $\mu \approx 10$ and $\omega/\Omega_{e0}=0.34$ at the equator. The estimated $\partial \omega/\partial t = 2.2\times 10^{-4}\,\Omega_{e0}^{-2}$ with $\alpha = 70\degree$. This is consistent with the measured chirping rate from the spectrogram at $z=0$ for $\omega/\Omega_{e0}=0.34$, which is $2.6\times 10^{-4}\,\Omega_{e0}^{-2}$.

Equation (\ref{eq:10}) from \citeA{Helliwell1967} has been shown to agree well with observations \cite{Tao2012b} in terms of the dependence of $\partial \omega/\partial t$ on radial distance. The value of $\partial\omega/\partial t$ directly from Helliwell's model is about a factor of 2 smaller than observations. However, as pointed out by \citeA{Tao2012b}, this is probably due to the working value of $\alpha$ being $30\degree$ from \citeA{Helliwell1967}, who assumed an isotropic pitch angle distribution and complete bunching for simplicity. If one takes into consideration that nonlinear wave particle interaction most easily occurs for high pitch angles \cite{Inan1978} and use, e.g., $\alpha= 72$\degree, the overall discrepancy between $\partial \omega/\partial t$ given by Equation (\ref{eq:10}) and observation is about $3\%$ \cite{Tao2012b}. Since Equation (\ref{eq:5}) of the TaRA model is the same as Equation (\ref{eq:10}) from Helliwell's model, and nonlinear wave interaction is taken into account ($\alpha\sim 70\degree$),  the results of \citeA{Tao2012b} also provide an observational support for the TaRA model.

In Helliwell's model, different tones are generated by shifting the interaction region to appropriate locations through the balance between input and output power. However, in PIC simulations, only rising-tone chorus has been reproduced so far with a dipole type background magnetic field. For a given inhomogeneity factor $\xi$, decreasing or increasing linear drive will only produce featureless weak or strong broadband emissions, respectively \cite{Katoh2013, Tao2014c, Tao2020}. Instead, spontaneous falling tone chorus has only been reproduced in PIC simulations with a reversed dipole magnetic field ($\xi < 0$), as demonstrated by \citeA{Wu2020b}. Correspondingly, generation of various forms of chirping elements through shifting the interaction region back and forth via power balance is not part of the TaRA model for spontaneous chorus. 

Other differences between the two models is about the extremely important role of nonlinear wave particle interactions, which result in $\partial\omega/\partial t \propto \delta B$ near the equator as illustrated in Figure \ref{fig:tar}. The nonlinear phase space dynamics of electrons also naturally explains the observed fine structures of chorus, such as chorus subpackets and the narrow bandwidth (see Section \ref{sec:fine-struct-chor}). Of course, the nonlinear wave particle interaction theory might not be well developed and the subpacket of chorus \cite{Santolik2004a} was probably not known in 1960s. 

\subsection{The sequential triggering model}
Another very influential model which explains chorus chirping is by \citeA{Omura2011}, called the sequential triggering model (see also \citeA{Katoh2013} and \citeA{Shoji2013}). In this model, the frequency chirping is due to the current parallel to the wave magnetic field, $\delta j_B$,  by
\begin{linenomath}
\begin{align}
  c^2k^2 - \omega^2 - \frac{\omega\omega_{pe}^2}{\Omega_e-\omega} = \mu_0 c^2k\frac{\delta j_B}{\delta B}.
\end{align}
\end{linenomath}
Note that, for consistency with \citeA{Omura2011}, SI units are used in this subsection, while the rest of the paper uses Gaussian-cgs units. By fixing $k$, a nonlinear resonant current $\delta j_B$ due to phase trapped electrons leads to a frequency increase $\delta \omega$ from the frequency of the triggering wave $\omega$. The triggered wave has frequency $\omega+\delta \omega$, and at the same time, a subpacket is formed. The new wave with frequency $\omega'=\omega+\delta\omega$ in turn plays the role of triggering wave, phase traps electrons forming $\delta j_B$, and changes frequency to $\omega'+\delta \omega'$. This triggering process occurs sequentially and forms a train of subpackets. Note here that $\delta \omega$ and $\delta \omega'$ are on the order of the frequency change over a subpacket, different from the infinitely small frequency spacing between modes ($\omega_i-\omega_{i-1}$) in the description of the TaRA model in Section \ref{sec:trap-release-tar}.

Both the TaRA model and the sequential triggering model emphasize the key role of nonlinear wave particle interactions in chorus chirping and share the same relation between $\partial \omega/\partial t$ and $\delta B$, Equation (\ref{eq:12}). However, there are three main differences between the two models. First, the most fundamental difference is about how chirping occurs. In the TaRA model, frequency chirping is due to the selective amplification of new emissions which satisfy the phase-locking condition to maximize power transfer at the release point in the upstream. This allows us to relate both the frequency chirping rate and its direction to the background field inhomogeneity, Equation (\ref{eq:5}). In the sequential triggering model, the chirping occurs due to $\delta j_B$, which is formed by nonlinear wave particle interactions near the equator \cite{Omura2011}. The chirping rate is related only to the wave amplitude as in Equation (\ref{eq:12}), a result from \citeA{Omura2008}, but not to the background magnetic field inhomogeneity. Second, in the sequential triggering model, each newly generated wave at $\omega'$ in turn plays the role of the triggering wave that phase-traps electrons to change the wave frequency until $\omega'+\delta \omega'$. In the TaRA model, all generated part of the chorus wave packet, from $\omega_0$ to $\omega_N$ as in Section \ref{sec:trap-release-tar}, could participate in nonlinearly phase-trapping new electrons to generate new emissions (Figure \ref{fig:tar}). Third, chorus subpacket is formed during the chirping process in the sequential triggering model. Each frequency sweeping from $\omega$ to $\omega+\delta \omega$ forms a subpacket. In this sense, chorus subpacket can be considered as the basic unit of wave excitation \cite{Katoh2013,Shoji2013,Hanzelka2020}. However, in the TaRA model, subpackets are formed by the conservation of momentum and energy between waves and phase-trapped electrons, see Section \ref{sec:fine-struct-chor} or \citeA{Tao2017a}. Therefore, chorus subpackets are more prominent in the downstream and its period could change with wave amplitude as waves are nonlinearly amplified while propagating downstream.

We note that \citeA{Shoji2013} applied the sequential triggering model to EMIC waves and took into consideration the generation of new emissions in the upstream, as observed in their PIC-type simulations. Following \citeA{Shoji2013}, \citeA{Hanzelka2020} developed a model of subpackets based on the sequential triggering model, which also has the feature of wave generation in the upstream. Unlike the TaRA model, however, the generated new emission is not due to the selective amplification through phase-locking at the release point, but due to $\delta j_B$ formed by nonlinear interactions with the previous triggering wave at the previous triggering point. Besides, both results of \citeA{Shoji2013} and \citeA{Hanzelka2020} are based on the sequential triggering model; therefore, comments above about the differences between the two models regarding how chirping and subpackets occur still apply.

\section{Fine structures of chorus}
\label{sec:fine-struct-chor}

\subsection{Chorus subpackets}
In the TaRA model, chorus subpackets are explained by conservation of momentum and energy between wave fields and phase-trapped electrons \cite{Tao2017a}. As phase-trapped electrons rotate in phase space, their velocity, and hence momentum, oscillates quasi periodically with period of $\mathcal{O}(\omega_{tr}^{-1})$. Conservation of momentum between waves and electrons naturally leads to that the wave amplitude oscillates at $\mathcal{O}(\omega_{tr}^{-1})$ \cite{Tao2017a,Tao2020,ONeil1965}. Based on this explanation of the subpackets, we can immediately arrive at a few conclusions as follows.

\begin{figure}
  \centering
  \includegraphics[width=0.8\textwidth]{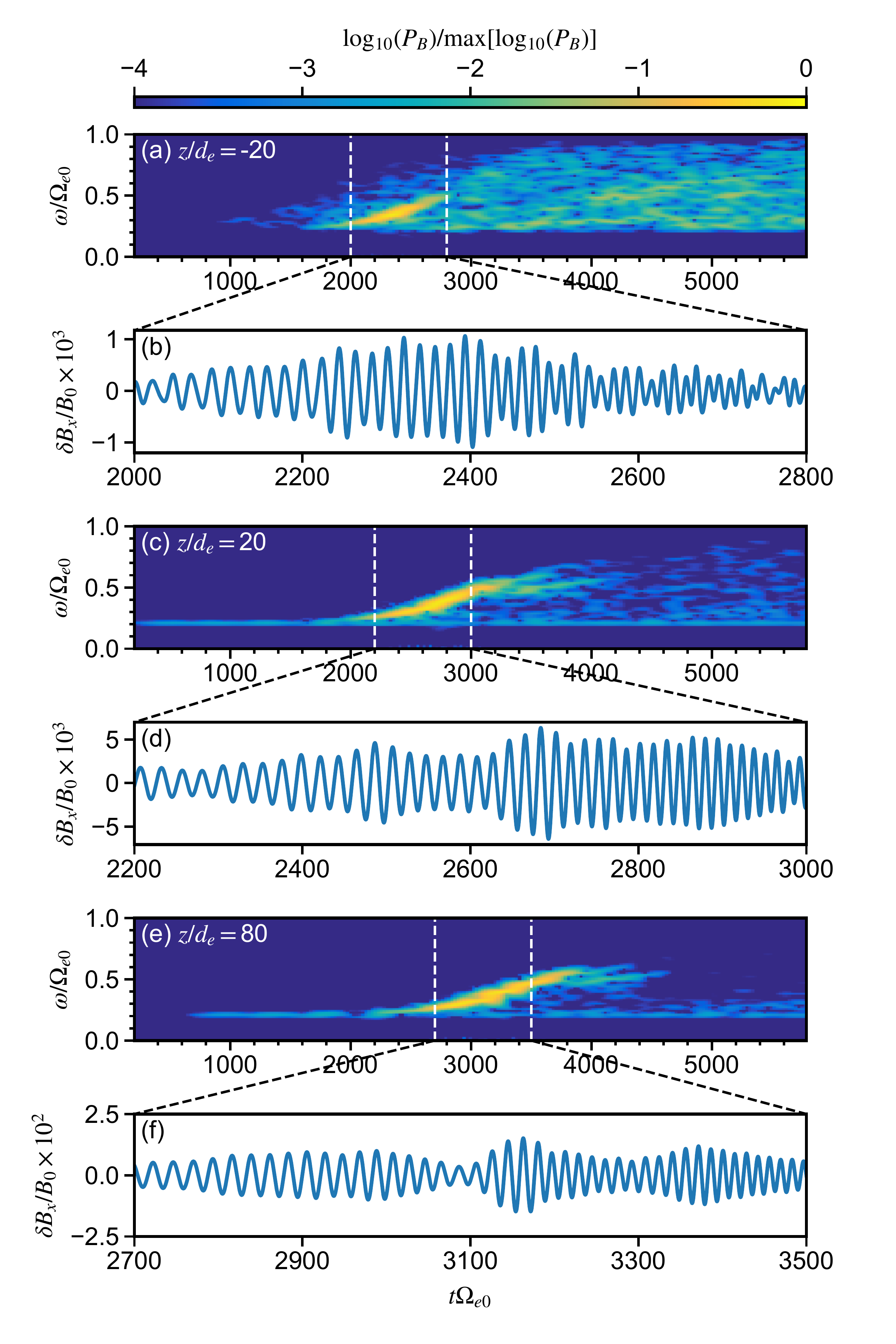}
  \caption{Comparison of chorus waveforms at $z/d_e=-20$, $20$, and $80$, together with corresponding wave spectrograms. In (a), (c), and (e), color-coded is the normalized wave power spectral density.}
  \label{fig:subpackets}
\end{figure}

First, in the TaRA model, there could be amplitude variation with frequency in the upstream due to the frequency dependence of the growth rate.  However, since phase trapping mainly occurs in the downstream, so are the quasi-periodic chorus subpackets. This is illustrated in Figure \ref{fig:subpackets}, where we show waveforms at three different locations: $z/d_e=-20, 20$, and $80$. We have bandpass filtered the waveform using Butterworth filter to remove the triggering wave at $0.2\Omega_{e0}$ . At $z/d_e=-20$, there is no obvious quasi-periodic variation of wave amplitude. However, in the downstream region, the subpackets are clearly visible at $z/d_e=20$ and even more so at $z/d_e=80$.  

Second, the number of subpackets within one element is not fixed, but could change as wave amplitude changes. Suppose the starting and ending frequencies of an element is $\omega_\text{i}$ and $\omega_\text{f}$, respectively, then the number of subpackets within one element, $N_s$, is
\begin{linenomath}
\begin{align}
  N_s \sim  (\omega_\text{f}-\omega_\text{i})\left(\pd{\omega}{t}\frac{2\pi}{\omega_{tr}}\right)^{-1}.
\end{align}
\end{linenomath}
Note that here we cannot use Equation (\ref{eq:12}) to arrive at the conclusion that $N_s$ is proportional to $\omega_{tr}^{-1}$ or $\delta B^{-1/2}$. This is because Equation (\ref{eq:12}) is valid only near the equator. Once generated, the frequency chirping rate does not change much (e.g., by dispersion) in the equatorial region. Correspondingly, we would arrive at $N_s\propto \omega_{tr} \propto \delta B^{1/2}$. This suggests that as waves propagate downstream and get amplified, the number of subpackets increases weakly with increasing $\delta B$, as long as the interaction remains nonlinear. This is not obvious in Figure \ref{fig:subpackets}, because the wave amplitude increases only by about a factor $2\sim 3$. However, this might provide a possible explanation for the observed short and intense subpackets within a chorus element \cite{Zhang2019,Zhang2018, Nunn2021}, which play important roles in nonlinear acceleration of radiation belt electrons \cite{Mourenas2018,Artemyev2020}. 

Third, the depth of modulation measured by $\delta B_\text{max}-\delta B_\text{min}$ is not fixed, but related to the magnitude of total momentum change during half the trapping period, $\Delta p$, of all phase-trapped particles. The  deeper wave amplitude modulation in the downstream is clearly consistent with the larger energy and momentum variation of phase-trapped electrons, as demonstrated in Figure \ref{fig:psi_change}a. The cause of this might be related to the larger wave amplitude in the downstream, as can be argued as follows. The exact calculation of $\Delta p$ is complicated in case of a fast chirping chorus element. But qualitatively, it could be expressed as
\begin{linenomath}
\begin{align}
  \Delta p \sim |\overline{\delta p}_+ N_+ - \overline{\delta p}_- N_-|, 
\end{align}
\end{linenomath}
where $\overline{\delta p}_+$ ($\overline{\delta p}_-$) is the average amount of momentum increase (decrease) per trapped particle during half the trapping period, and $N_+$ and $N_-$ are the related number of particles. For simplicity of discussion, we assume $\overline{\delta p}_+ = \overline{\delta p}_- = \overline{\delta p}$; therefore,
\begin{linenomath}
\begin{align}
  \label{eq:16}
  \Delta p \sim |\overline{\delta p} \times \Delta N|, 
\end{align}
\end{linenomath}
where $\Delta N \equiv N_+ - N_-$, which is clearly related to the phase space density gradient at the resonant velocity. If, for example, after sufficiently phase-mixing such that the phase-space density within the trapping region is constant, then $\Delta N = 0$ and there is no amplitude modulation. On the other hand, $\delta p$ is related to the width of the resonant island, which is $\propto \omega_{tr}/k$ if measured in terms of $v_\|$. Therefore, a larger wave amplitude leads to a larger $\delta p$, and correspondingly a deeper amplitude modulation, if other parameters are fixed. This crude argument suggests that the increasing wave amplitude might contribute to the deeper amplitude modulation at $z/d_e=80$ than at $z/d_e=20$ in Figure \ref{fig:subpackets}.  Of course, a more quantitative calculation is needed to confirm this argument. 

Fourth, although the subpacket period $T$ is estimated to be on the same order as $\omega_{tr}^{-1}$, we may provide a refined estimate of $T$ by taking into account that the resonant electrons move oppositely to the wave. Suppose the observation of the waveform and subpackets is made at a location $z_0$, where the wave amplitude maximizes at time $t_0$ due to momentum exchange with electrons. The same group of phase-trapped electrons would cause another wave amplitude maximum after $T_{tr}$, which is $2\pi/\omega_{tr}$, at $z_0 + \Delta z$ with $\Delta z = v_r T_{tr}$. This new maximum will propagate to $z_0$ after a duration of $\Delta z/v_g$. Correspondingly, for an observer at $z_0$, the time difference $T$ between two maximum is about
\begin{linenomath}
\begin{align}
  T \sim \left(1-\frac{v_r}{v_g}\right)T_{tr}. 
\end{align}
\end{linenomath}
Noting $v_r/v_g < 0$, $T/T_{tr} > 1$. For typical cases with $\mu \gg 1$, $v_r/v_g = -(2\omega/\Omega_e)^{-1}$. Therefore, $T\sim 3 T_{tr}$ if $\omega/\Omega_e = 0.3$, or $2 T_{tr}$ if $\omega/\Omega_e = 0.5$. For the waveform shown in $z/d_e = 80$, we may take $\Omega_e \approx \Omega_{e0}$, $\delta B/B_0 \sim 10^{-2}$, and therefore, $\omega_{tr}/\Omega_{e0}  \sim 0.1$ with $\mu \sim 10, v_\perp/c \sim 0.3, \omega/\Omega_{e0}\sim 0.3$. This leads to $T_{tr} \sim 60$, and $T\sim 180$. The period roughly agrees with the period of subpacket at $z/d_e=80$ shown in Figure \ref{fig:subpackets}f.

In summary, we conclude here that the conservation of momentum and energy between phase-trapped electrons and chorus waves is responsible for both the formation and the dynamic evolution of chorus subpackets. A quantitative description of subpackets should take into account the conservation laws and dynamics of phase-trapped electrons. 

\subsection{Instantaneous bandwidth of chorus elements}
\begin{figure}
  \centering
  \includegraphics[width=\textwidth]{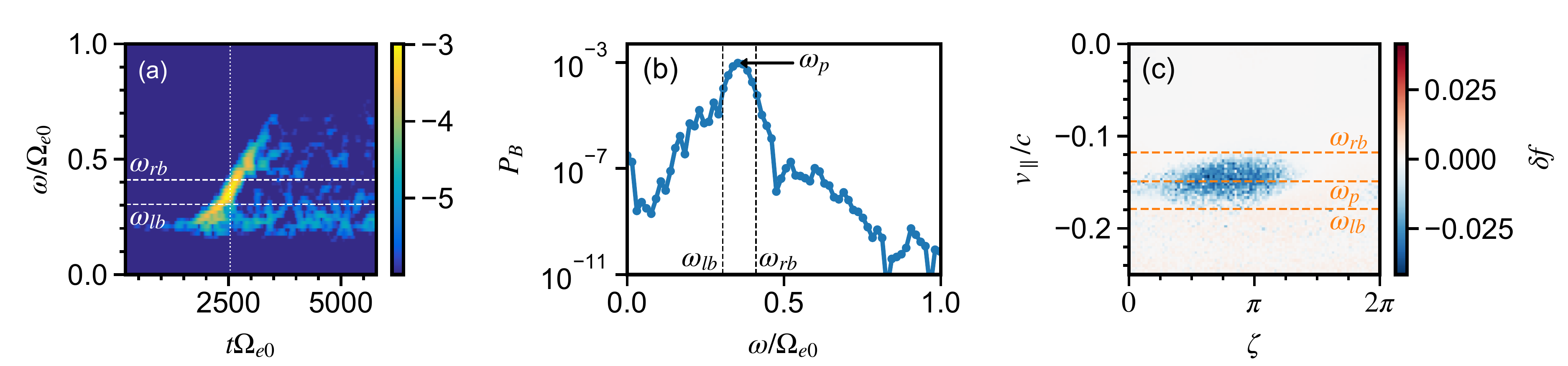}
  \caption{Estimation of the instantaneous bandwidth of chorus. (a) The wave spectrogram at $z/d_e=0$. (b) The power spectral density as a function of frequency at the time marked by the vertical dotted line in Panel (a). (c) The electron phase space distribution. In all three panels, $\omega_{lb}$ and $\omega_{rb}$ are the left and right boundaries marking locations of full width half maximum of $\log_{10}(P_B)$ in Panel (b), and $\omega_p$ is the frequency of peaking wave intensity.}
  \label{fig:ibandwidth}
\end{figure}

Chorus waves are narrowband emissions, and the instantaneous bandwidth should be on the order of $\omega_{tr}$ due to nonlinear wave particle interactions, as illustrated in Figure \ref{fig:ibandwidth}. We use the wave spectrogram at $z=0$ as an example (Figure \ref{fig:ibandwidth}a). To estimate the bandwidth, we plot the wave power spectral density $P_B$ as a function of frequency at $t\Omega_{e0} = 2537$, as indicated by the vertical dashed line in Figure \ref{fig:ibandwidth}a. This particular time is chosen simply because there is a saved distribution of electrons. In Figure \ref{fig:ibandwidth}b, we estimate the bandwidth $\Delta \omega$ by the full width at half maximum (FWHM) of $\log_{10}(P_B)$. The left and right boundaries are denoted as $\omega_{lb} \approx 0.3\Omega_{e0}$ and $\omega_{rb} \approx 0.4\Omega_{e0}$; therefore, $\Delta \omega \approx  0.1\Omega_{e0}$. From Figure \ref{fig:ibandwidth}c, the two frequencies $\omega_{lb}$ and $\omega_{rb}$ roughly correspond to the resonant frequencies at the lower and upper boundaries of the resonant island. Since the width of the resonant island is $\mathcal{O}(\omega_{tr}/k)$ measured by $v_\|$, the difference between $\omega_{lb}$ and $\omega_{rb}$ or the instantaneous bandwidth of chorus is $\mathcal{O}(\omega_{tr})$. A more accurate estimate, of course, needs to take into account the value of $R$; e.g., if $R = 0$, the resonant island width is $4\omega_{tr}/k$. In case of Figure \ref{fig:R}, $R\approx 0.5$; therefore, $\Delta v_\| \approx 2 \omega_{tr}/k$. From Equation (\ref{eq:1}) and $\delta B/B_0 \approx 2\times 10^{-3}$ at $z=0$, we may estimate $\omega_{tr}/\Omega_{e0} \approx 0.045$. Neglecting change of $k$, the resonant island width indicates that the bandwidth of chorus should be $0.09\Omega_{e0}$, consistent with the estimate by $|\omega_{lb}-\omega_{rb}|$. 

We may use typical parameters in the magnetosphere to estimate the order of instantaneous bandwidth of chorus elements. We choose $\omega/\Omega_{e0} = 0.3$ as a representative frequency. Outside the plasmapause, we may take $\mu \sim 10$, and therefore $v_r \approx 0.2c$. With $v_\perp \sim 0.5c$, which corresponds to $\alpha \approx 70\degree$, we arrive at $\omega_{tr}/\Omega_{e0} \approx (\delta B/B_0)^{1/2}$ from Equation (\ref{eq:1}). At $L=5$, $B_0\approx  0.0024$\,G, and $\Omega_{e0} \approx 42240$\,rad/s. For the typical amplitude $\delta B/B_0\sim 10^{-3}$, the bandwidth is on the order of $1335$\,rad/s or $212$\,Hz. This roughly agrees with observation that chorus elements typically have a bandwidth of a few hundred Hz.

Note that the bandwidth of chorus has a dependence on wave amplitude as $\delta B^{1/2}$. Correspondingly, stronger (weaker) elements tend to have larger (smaller) instantaneous bandwidth. 

\section{Summary}
\label{sec:summary}
In this work, we presented a theory-based phenomenological model of chorus, the TaRA model, based on analysis of effective wave growth rate and electron phase space dynamics from a PIC simulation with the DAWN code. The TaRA model shares the same principle with many previous models in that nonlinear wave particle interactions phase organizes energetic electrons, and these phase-correlated electrons generate new coherent emissions. The difference among various models is on how the principle is applied to explain chirping and other properties of chorus in details. Our results could be summarized as follows:

1, The effective wave growth rate $\gamma_\text{eff}$ is much larger in the upstream than in the downstream. This, combined with electron phase space dynamics analysis, demonstrates that the upstream region is mainly for generation of new emissions and wave amplification, while the downstream is mainly for phase-organizing resonant electrons by nonlinear wave particle interactions. The fact that $\gamma_\text{eff} \gg \gamma_L$ at wave excitation ensures the generation of narrowband emissions by phase-correlated electrons.   

2, The TaRA model explains chirping of chorus by selective amplification of new emissions that satisfy the phase-locking condition with phase-correlated electrons released from the previously generated part of the packet. The phase-locking condition allows the longest possible resonant interaction and maximizes possible wave power transfer. At the release point, this condition leads to that chorus frequency chirping rate is determined by the background magnetic field inhomogeneity for the rising-tone chorus analyzed in this study. This chirping rate has the same form as the one given by \citeA{Helliwell1967}. 

3, The nonlinear wave particle interaction mainly occurs in the downstream. Correspondingly, the relation between frequency chirping rate and wave amplitude originally proposed by \citeA{Vomvoridis1982} is also valid near the equator. The validity of this relation could be regarded as a proof of the important role played by nonlinear wave particle interactions in generation of chorus. 

4, The TaRA model, therefore, unifies two different ways of estimating chorus chirping rate from \citeA{Helliwell1967} and \citeA{Vomvoridis1982}, and suggests that the difference between the two chirping rates could be explained by that they are derived at different stages of chorus generation. Correspondingly, this model explains why both chirping rates have been shown to agree very well with observations and simulations in previous studies. 

5, Subpackets of chorus are explained by the conservation of momentum and energy between wave fields and phase-trapped electrons as in \citeA{Tao2017a}. This suggests that subpacket period and the number of subpackets for a given element are not fixed, but have a week dependence on wave amplitude, exhibiting dynamic evolution as waves propagate from upstream to downstream.

6, The bandwidth of chorus is on the order of $\omega_{tr}$, which suggests that stronger elements tend to have larger bandwidth than weaker elements. 

Our model is based on a PIC simulation for rising-tone chorus with a dipole-type background magnetic field. By properly taking into account different background plasma parameters and resonance conditions, it is potentially possible to explain various chirping phenomena previously reported with the same principle, including chirping of chorus in a uniform field, falling tone chorus, and chirping of electromagnetic ion cyclotron waves in the magnetosphere. The quantitative results obtained in the study should also be helpful to understand different properties of chorus observed at Earth, Saturn \cite{Hospodarsky2008} and Jupiter \cite{Menietti2008c}. Finally, we emphasize that the purpose of the phenomenological TaRA model is mainly to elucidate how and where chirping occurs. A more quantitative analysis could be based on the theoretical framework of \citeA{Zonca2017} and \citeA{Zonca2021}.  

\section{Acknowledgments}
This work was supported by NSFC grants (41631071 and 11235009), the Strategic Priority Program of the Chinese Academy of Sciences (No.~XDB41000000), the Fundamental Research Funds for the Central Universities, EURATOM research and training programme 2014-2018 and 2019-2020 under grant agreement No.~633053. The views and opinions expressed herein do not necessarily reflect those of the European Commission. Code and input files used for generating the data used in this study can be found at \url{https://doi.org/10.5281/zenodo.4774932}.


\bibliography{xt}

\end{document}